\documentclass[12pt]{article}
\usepackage{amsfonts}
\usepackage{amsthm}
\usepackage{fancyhdr}
\usepackage{comment}
\usepackage{abraces}
\usepackage{times}
\usepackage{amsmath}
\usepackage[margin=1in]{geometry}
\usepackage{amssymb}
\usepackage{epsfig}
\usepackage{mathtools}
\usepackage{multirow}
\usepackage{natbib}

\newtheorem{corollary}{Corollary}

\newtheorem{definition}{Definition}

\newtheorem{observation}{Observation}

\newtheorem{proposition}{Proposition}
\newtheorem{remark}{Remark}

\newtheorem{assumption}{Assumption}
\DeclareMathOperator*{\argmax}{arg\,max}

\title{Algorithms, Incentives, and Democracy\thanks{Comments welcome at  \textit{elizabeth.m.penn@gmail.com} and/or \textit{jwpatty@gmail.com}. We thank Avi Acharaya, Steve Callander, John Duggan, Sandy Gordon, Cathy Hafer, Alex Hirsch, Zhaotian Luo, Alastair Smith, Randy Stevenson, and Rick Wilson for incredibly helpful comments and suggestions. All remaining errors are our own.} 
}
\author{Elizabeth Maggie Penn (Emory) \and John Patty (Emory)}
\date{\today}

\begin{document}
\bibliographystyle{apsr}
\maketitle
\begin{abstract}

Classification algorithms are increasingly used in areas such as housing, credit, and law enforcement in order to make decisions affecting peoples' lives. These algorithms can change individual behavior deliberately (a fraud prediction algorithm deterring fraud) or inadvertently (content sorting algorithms spreading misinformation), and they are increasingly facing public scrutiny and regulation.  Some of these regulations, like the elimination of cash bail in some states, have focused on \textit{lowering the stakes of certain classifications}.  In this paper we characterize how optimal classification by an algorithm designer can affect the distribution of behavior in a population---sometimes in surprising ways.  We then look at the effect of democratizing the rewards and punishments, or stakes, to algorithmic classification to consider how a society can potentially stem (or facilitate!) predatory classification.  Our results speak to questions of algorithmic fairness in settings where behavior and algorithms are interdependent, and where typical measures of fairness focusing on statistical accuracy across groups may not be appropriate.

\end{abstract}

\section{Introduction}

Algorithms are ubiquitous in modern life, particularly \textbf{classification algorithms}, which assign individuals, texts, pictures, and/or other things to \textit{categories}.  High profile examples of such algorithms include credit scores, the COMPAS scoring algorithm intended to predict recidivism risk, and facial recognition systems.  When these algorithms classify behavior, they may also affect behavior. An algorithm designed to trigger an audit when fraud is suspected may also serve to reduce the level of fraud that people engage in.  An algorithm designed to evaluate college readiness may also promote college readiness in a population of students. These examples illustrate an important aspect of classification algorithms: not only are they are used to make decisions that affect peoples' lives, but they also affect the life choices that people make.

Over the past 25 years there has been increasing attention paid to the proper usage and design of classification algorithms.  For example, some governments have regulated the use of various types of algorithmic classification data for some decisions within the realm of credit, housing, and employment.  These are not bans on the algorithms themselves, of course --- rather these regulations are more properly thought of as \textit{reducing the stakes} of some algorithmic classifications.  A high-profile example is the elimination of cash bail, which reduces the stakes of receiving a high pretrial release risk score. Another example is the prohibition of credit scores to determine a person's eligibility for housing, which reduces the stakes of classification on the basis of creditworthiness. While algorithms could of course (at least in theory) be directly regulated, there are many reasons that democratic intervention regarding classification algorithms tends to be focused on the stakes of the algorithm's determinations, rather than the details of the algorithm itself.\footnote{There are some exceptions to this.  For example, some data is subject to privacy protections, algorithms can be audited and/or subject to minimum accuracy/reliability thresholds, and so forth.  But, aside from situations in which the algorithm is being used within a specific field (such as health care) or by a regulated entity (including government agencies themselves), such ``statistical'' intervention is rare in practice.}  

We present a model motivated by these two facts: algorithms affect behavior in potentially meaningful ways, and the stakes of an algorithm may be an object of democratic choice.  Specifically, we consider a situation in which individuals in a population make a binary, and potentially costly, choice about whether to \textit{comply} or \textit{not}.  There exists a \textbf{designer} with preferences over both compliance and over how individuals are rewarded and punished.  The designer's algorithm will classify each individual as either a ``1" (deserving of a reward) or as a ``0" (deserving of a penalty), conditional on a noisy signal of each individual's choice.  The designer can implement any classification algorithm he wants in order to further his goals. 

We begin by considering the optimal classification algorithm for the designer when the stakes of the algorithm (the rewards or punishments meted to individuals) are exogenous. In this setting, the designer's objectives will play a large role in the equilibrium rates of compliance that are observed. Perhaps unsurprisingly, with a sufficiently punitive system of rewards and penalties, a designer can induce anywhere from virtually 0\% compliance to virtually 100\% compliance, simply through his choice of classification algorithm.  Moreover, with a sufficiently punitive system of rewards and penalties  the designer can induce a distribution of classification outcomes that are virtually 100\% true positives (rewarded compliers); virtually 100\% true negatives (penalized non-compliers);  virtually 100\% false negatives (penalized compliers); or virtually 100\% false positives (rewarded non-compliers).

We then consider what a designer can accomplish when the rewards and punishments stemming from classification are chosen democratically by the individuals who will be classified by the algorithm.  We assume that these rewards and penalties satisfy a budget balance condition, so that the net penalties paid by individuals classified as a ``0" are redistributed to the individuals classified as a ``1."  Individuals face varying personal costs to compliance, and may also have a taste for aggregate compliance.  In equilibrium, the optimal classifier and the optimal system of rewards are mutually reinforcing.  More specifically, the optimal classifier represents a best response by the designer to the democratically chosen system of rewards and penalties. Moreover, conditional on the optimal classifier chosen by the designer, the democratically chosen system of rewards is a Condorcet winner.

Perhaps the most powerful finding from the model is that democratic institutions not only constrain the algorithm designer --- in many cases, they will totally circumscribe the kinds of behaviors the designer can incentivize.  Our results can be cast in terms of two possible types of classifiers. The first we term a \textit{null classifier}; this type of classifier disregards signal information about individuals' compliance decisions, and classifies every individual as deserving of reward or penalty with equal probability. In the presence of a null classifier, individuals only comply if they have an intrinsic taste for compliance. We term any classifier that utilizes signal information as \textit{non-null}.  So long as rewards and penalties are differentiated, non-null classifiers always induce some individuals to alter their behavior relative to their intrinsic taste for compliance.

Surprisingly, we show that when rewards and punishments are democratically chosen, \textit{every non-null classifier induces the same level of equilibrium compliance}.  This occurs because the median voter's preferences for rewards and penalties reduce to a preference for an optimal aggregate level of compliance.  For any non-null classifier, a system of rewards and penalties can always be designed to induce the median's optimal level of compliance.  Consequently, the designer's preferences can have no effect on equilibrium compliance in any equilibrium in which the optimal classifier is non-null.  However, both the designer and the voters can always induce a ``null" outcome in which only individuals with an intrinsic taste for compliance comply.  The designer can achieve this through any null classifier, while the voters can achieve this by setting the rewards and penalties associated with compliance equal to zero.

In some instances these types of null equilibria are the only possible equilibria, and this can be socially desirable. When the preferences of voters and the designer are at odds, in terms of their taste for aggregate compliance, it may be the case that no non-null equilibrium is attainable.  If the designer, for example, seeks to maximize ticket revenue by using an algorithm that incentivizes non-compliance, and if the median voter values aggregate compliance, then democratic institutions can serve to disable the (potentially predatory) algorithm.  On the other hand, there may also be instances in which the democratic choice of rewards and punishments leads to a unique null equilibrium that represents an inferior outcome for the median voter, for the designer, and for aggregate social welfare. In this case, it may be better to take the decision to set rewards and penalties away from the public.  Finally, we can construct examples, similar to a game of matching pennies, in which there are no (pure strategy) equilibria.

\subsection{Related Literature}

Our theory draws from a long-running literature on the political economy of public policy, the emerging literature on \textit{algorithmic fairness} (or algorithmic bias), and a recent literature on what we refer to as \textbf{algorithmic endogeneity}.  

\paragraph{Political Economy.}  The relevant literature on political economy is rich and largely well-known.  Our model ultimately employs a version of the seminal framework developed by \cite{MeltzerRichard81} to consider how democratic choice and investment incentives interact in equilibrium.  Accordingly, some of our results mirror theirs (\textit{e.g.}, individuals do not fully internalize the social benefits of public policy, and democratic choice is generally socially inefficient).  As in their model, the democratic process in equilibrium is essentially equivalent to the preferences of the median voter (\cite{Black48}).   Additionally, our model introduces the beginnings of a \textit{principal-agent problem} (\cite{Gailmard14}).  While we do not consider selection or retention in our model, the model allows for the algorithm designer (the agent) to have different preferences than the voters (the ``principals'') and illustrates how such divergence can affect public policy in equilibrium.  In particular, when the preferences of the principal and agent are sufficiently opposed to each other, public policy will be completely ineffective in equilibrium.

The most closely related paper in this vein is the recent contribution by \cite{Alexander22}, who considers the collective preferences over using carrots (positive rewards) or sticks (negative penalties) to induce socially desirable behavior.  Alexander's analysis illustrates that --- within a Meltzer-Richard style framework --- carrots and sticks are not equivalent in democratic choice environments, because each voter should take into account his or her own likelihood of receiving a reward versus benefiting from others being fined.  Our analysis complements Alexander's, particularly in terms of identifying an implicitly ``predatory'' motivation on the part of the median voter when choosing the reward or penalty that the algorithm will impose on all individualls in the population.

\paragraph{Algorithmic Fairness.}  Beginning about 15 years ago, scholars and policymakers began to focus on how algorithms, and the data on which they are based, might treat people unfairly (\textit{e.g.}, \cite{DworkHardtPitassiReingoldZemel12}).  This emergence followed a series of findings across multiple policy areas that demonstrated that racial, gender, or other forms of bias often characterize algorithmic decision-making in important policy area.  A high-profile example of this is housing and lending (\textit{e.g.}, \cite{Ladd98}, \cite{MunnellTootellBrowneMcEneaney96}, \cite{FoggoVillasenor21}).  It has also been documented in criminal justice (\textit{e.g.}, \cite{Angwin2016} and \cite{Washington18}), college admissions (\textit{e.g.}, \cite{KleinbergLudwigMullainathanRambachan18}, \cite{MartinezNedaZengGagoMasague21}), and advertising (\textit{e.g.}, \cite{MillerHosanagar19}).  The range of these findings, along with early theoretical results (\textit{e.g.}, \cite{KleinbergMullainathanRaghavan16}, \cite{Chouldechova17}) prompted scholars to develop and compare different notions of fairness in algorithmic settings.\footnote{For some early discussions of the different notions of algorithmic fairness, see \cite{KusnerLoftusRussellSilva17}, \cite{CorbettDaviesGoel18}, \cite{Narayanan18}, \cite{ChouldechovaRoth18}, \cite{BerkHeidariJabbariKearnsRoth18}, \cite{KleinbergMullainathan19}, and \cite{Sharifi19}. For a recent overview, see \cite{MitchellPotashBarocasDAmourLum21}. }  Unsurprisingly, the relationship between algorithmic fairness and economic theories of discrimination was soon noted (\textit{e.g.}, \cite{KleinbergEtAl18}, \cite{LangSpitzer20}, and \cite{PattyPenn21PhilCompass}).\footnote{Several earlier works on employment discrimination presaged the current state of the literature, including \cite{CoateLoury93}, \cite{FangMoro11}, and \cite{Fryer07}.}  
While our focus in this article is not algorithmic fairness, per se, our results extend this literature by considering the consequences of allowing individuals subject to an algorithm to have a role in determining the impact of the algorithm on all individuals subject to it.\footnote{A recent consideration of a setting similar to ours that is more squarely focused on algorithmic fairness is offered by \cite{PattyPenn23AlgorithmicEndogeneityFairness}.}

\paragraph{Algorithmic Endogeneity.}  Our model allows both the algorithm designer and the voters to play a role in choosing the algorithm and, accordingly, endows all of the actors with preferences over the algorithm's decisions and the data (i.e. the individual behaviors) the algorithm induces in equilibrium.  This brings our results into conversation with the very new literature on what we term \textit{algorithmic endogeneity}: the algorithm affects the data distribution, and the data distribution affects what one considers an optimal algorithm.  Interest in this topic truly emerged a little less than a decade ago.  \cite{HardtEtAl16} defined the notion of \textit{strategic classification}, which captures the notion of an optimal algorithm when the algorithm affects the data distribution itself.  This concept was subsequently generalized under the moniker \textit{performative prediction} (\cite{PerdomoEtAl20}).  Our results augment this early literature primarily through its focus on strategic individual choice as the foundation of how changes in the algorithm affect the data distribution itself.  As others have noted (\textit{e.g.}, \cite{KleinbergEtAl18}, \cite{PattyPenn23PerfectPrediction}), this step is required before one can judge any algorithm's \textit{welfare impact}. Our model is also related the ``manipulation'' of algorithms through data manipulation (\cite{FrankelKartik22}).

%
%\cite{KleinbergMullainathan19}; \cite{Sharifi19}; \cite{Lang20}; \cite{Becker71}; \cite{MunnellTootellBrowneMcEneaney96}; \cite{Ladd98}; \cite{FoggoVillasenor21};
%\cite{Washington18}; \cite{MillerHosanagar19}; \cite{MitchellPotashBarocasDAmourLum21}; \cite{Narayanan18}
%\cite{CorbettDaviesGoel18}; \cite{BerkHeidariJabbariKearnsRoth18}; 
%\cite{Angwin2016}; \cite{Fryer07}; \cite{KusnerLoftusRussellSilva17}; \cite{LiptonChouldechovaMcAuley18}; \cite{Fang11}; \cite{CorbettDaviesEtAl17}
%\cite{ChouldechovaRoth18}; \cite{Jung20}; \cite{FrankelKartik22}

\section{The Model}

We consider a model of individual behavior and algorithmic classification, in which a continuum of \textbf{individuals}, $N=[0,1]$, is faced with choosing from a set of two \textbf{behaviors}, $B=\{0,1\}$.  After making his or her individual choice of behavior, $\beta_i \in B$, each individual $i \in N$ will be assigned a \textbf{decision}, $d_i\in \{0,1\}$,  by another actor, referred to as the \textbf{algorithm designer}, who is denoted by $D$. The designer $D$ makes this assignment choice for each individual on the basis of potentially noisy information about $\beta_i$, and the designer, $D$, may have preferences over both $\beta \equiv \{\beta_i\}_{i \in N}$ and $d \equiv \{d_i\}_{i \in N}$.  

With the noisiness of his or her information about individual behaviors and his or her preferences in hand, $D$ must design a \textbf{classification algorithm} (or \textbf{classifier}) that rewards or punishes individuals for certain types of behaviors. The classification algorithm will map a noisy but informative \textbf{signal} received regarding individual $i$'s behavior, $s_i \in \{0,1\}$, into a decision, $d_i \in \{0,1\}$.  Each individual $i \in N$ will observe the details of the algorithm and choose their behavior, $\beta_i \in B$, possibly incurring a cost to affect the algorithm's decision regarding $i$, or $d_i$.

\subsection{Timing of Decisions}

At the beginning of the game, each individual $i \in N$ privately observes his or her \textbf{type}, $\gamma_i$, which represents his or her net cost of choosing $\beta_i=1$ (as opposed to $\beta_i=0$).  Note that $\gamma_i$ can be negative, which implies that individual $i$ experiences a direct net \emph{benefit} from choosing $\beta_i=1$.\footnote{For presentational simplicity, we normalize the continuum of individuals by ordering the individuals according to their types: $i < j \Leftrightarrow \gamma_i \leq \gamma_j$.}

\begin{remark}
    If we stopped here and omitted the algorithm designer (and algorithm) from the analysis, each individual's optimal choice of $\beta_i$ is simply
    \begin{equation}
        \label{Eq:SincereBehavior}
        \beta_i^S(\gamma_i) = \begin{cases}
        1 & \text{ if } \gamma_i\leq 0,\\
        0 & \text{ if } \gamma_i> 0.
        \end{cases}
    \end{equation}
    This represents \textit{sincere behavior}, uncontaminated by incentives emanating from individuals' preferences over the decision rendered by the algorithm. We refer to this behavioral baseline later when characterizing the effects of the designer's preferences and algorithm design choices.
\end{remark}

\paragraph{Type Distribution.}  We assume that there is a unit mass of individuals whose costs are distributed according to a cumulative distribution function (CDF) denoted by $F: \mathbf{R} \to [0,1]$, and that $F$ is continuously differentiable, with probability density function (PDF), denoted by $f: \mathbf{R} \to \mathbf{R}_{+}$.  In addition, we make the following assumption about the PDF, $f$.

\begin{assumption}\label{cdfAss} 
Individuals' costs are distributed according to a log-concave PDF $f$ with full support on $\mathbf{R}$.% ($f(\gamma_i)>0$ for all $\gamma_i \in \mathbf{R}$).
\end{assumption}

The assumption that $f$ possesses full support implies that, in equilibrium, any algorithm designed by $D$ will induce a positive mass of individuals to choose $\beta_i=0$ and a positive mass to choose $\beta_i=1$.  For reasons that will become clear, we refer to $F(0)$ as the level of \textbf{sincere prevalence}.

\paragraph{The Designer's Problem.}  Simultaneous to the individuals' observing their types, the designer, $D$, designs an algorithm, $\delta \equiv (\delta_1,\delta_0) \in [0,1]^2$.  An algorithm is a pair of probabilities, each representing the probability that any individual will be classified according to their signal:
\[
\Pr[d_i = s_i \mid \delta] = \delta_{s_i}.
\]

\paragraph{The Signal Structure.}   The inputs to the algorithm (the binary signal for each individual $i$) are signals that are noisy, but correlated with the individual's choice of behavior, $\beta_i$.  After observing $\delta$ and $\gamma_i$, each individual $i$ simultaneously chooses a behavior, $\beta_i \in B$.  Conditional on $\beta_i$, the signal $s_i \in \{0,1\}$ is generated according to the following distribution:
\[
\Pr[s_i = \beta_i] \equiv \phi \in \left(\frac{1}{2},1\right].  
\]
The probability $\phi$ represents the accuracy of the information about $\beta_i$ for each individual $i \in N$,\footnote{We assume that, conditional on $(\beta_i, \beta_j)$ for any distinct pair of individual $i\neq j \in N$, $s_i$ is distributed independently across individuals.} and is assumed to be common knowledge throughout.  When $\phi=1$, the algorithm is capable of 100\% accuracy in rendering its decisions.  This extreme baseline will be helpful from time to time as we illustrate the origins of the incentives identified by our analysis below.

\subsection{Individual Preferences} 

Each individual $i \in N$'s payoff function is as follows:
\begin{equation}
    \label{Eq:BaselineVoterPayoff}
    u_i(\beta_i,d_i \mid \gamma_i, r, t) = r \cdot d_i - \beta_i \cdot \gamma_i.
\end{equation}
%The first term on the right hand side of \eqref{Eq:BaselineVoterPayoff} represents the proportion of people complying ($\beta_j=1$) in the population (the integration from 0 to 1 is with respect to the citizens' indices: $N=[0,1]$).  The marginal value of compliance to any given voter $i$ is $t\geq 0$; $t$ captures the degree to which individuals value aggregate compliance (an example would be public safety).\footnote{Note that, while the integral in Equation \eqref{Eq:BaselineVoterPayoff} might not exist in general, it will always be well-defined in our setting if the distribution of costs is absolutely continuous with respect to Lebesgue measure and individuals choose optimally.} 
The term $r$ captures an exogenous \textbf{reward} to classification (for expositional clarity, we'll refer to $r$ as a ``reward" even when $r$ is negative and represents a penalty).  Each voter $i \in N$ classified as $d_i = 1$ by the algorithm receives an additive payoff of $r \in \mathbf{R}$, and no additional payoff if $d_i=0$.

%\begin{remark}
%    We will augment the payoff function in Equation \eqref{Eq:BaselineVoterPayoff} when we endogenize the reward, $r$. Also, note that $r$ is a net reward for being classified as a $d_i=1$ as opposed to being classified as $d_i=0$.  We will return to this point again when we endogenize the net reward, $r$.
%\end{remark}

Note at this point that, if $r=0$, the sincere strategy, $\beta_i^S$, as defined in Equation \eqref{Eq:SincereBehavior}, is optimal.  However, if $r \neq 0$ then, generically, some individuals will find a different strategy optimal.\footnote{Here, the genericity is with respect to Lebesgue measure on $\mathbf{R} \times [0,1]^2$.}  We will of course return to this below.  Prior to that, we complete the setup of the model by describing the algorithm designer's preferences.

\subsection{The Designer's Preferences} 

Our designer's preferences are potentially over either (or both) individual behavior and the accuracy of the algorithm's determinations.  Specifically, we assume that $D$'s payoff from the algorithm assigning decision $d_i\in \{0,1\}$ to an individual $i$ who chose behavior $\beta_i\in \{0,1\}$ is equal to
\begin{equation}\label{dUtility}
u_D (d_i,\beta_i) \equiv \begin{cases}
A_1 & \text{ if $\beta_i=1$ is rewarded ($d_i=1$)},\\
A_0 & \text{ if $\beta_i=1$ is not rewarded ($d_i=0$)},\\
B_1 & \text{ if $\beta_i=0$ is rewarded ($d_i=1$), and}\\
B_0 & \text{ if $\beta_i=0$ is not rewarded ($d_i=0$)}.
\end{cases}
\end{equation} 
with $(A_0,A_1,B_0,B_1) \in \mathbf{R}_+$ being exogenous and commonly known.\footnote{The assumption that $A_1, A_0, B_1, B_0\geq 0$ is without loss of generality, as $D$'s behavior is unique up to a positive affine scaling of these payoffs. This would change qualitatively if we consider $D$'s incentive to invest in increasing the accuracy of the signal, $\phi$.}  To save space, we will denote the designer's preferences by $\eta \equiv \{A_1,A_0,B_1,B_0\}$. 
Rewriting \eqref{dUtility} as a function of the algorithm's confusion matrix, the designer's ex post payoff, conditional on $\beta_i$ and $d_i$, is defined by the following:
\begin{table}[hbtp]
    \centering
    \begin{tabular}{|c||c|c|} \hline
&\multicolumn{2}{c|}{Decision} \\ \hline
Behavior&$d_i=1$&$d_i=0$ \\ \hline \hline

\multirow{2}{*}{$\beta_i=1$}&$A_1$&$A_0$ \\ 
& (True Positive)&(False Negative) \\ \hline
\multirow{2}{*}{$\beta_i=0$}&$B_0$&$B_1$\\ 
& (False Positive) & (True Negative) \\ \hline
 \end{tabular}
    \caption{The Designer's Ex Post Payoffs}
    \label{Tab:DesignerExPostPayoffs}
\end{table}

\noindent Table \ref{Tab:DesignerExPostPayoffs} will be useful in carrying out, and interpreting, our analysis that follows.  Now we turn to consider how these payoffs shape $D$'s incentives when designing the algorithm.

\subsection{The Algorithm Designer's Objectives\label{Sec:Objectives}}

The functional form provided by Equation \ref{dUtility} can capture a variety of objectives of the algorithm designer, and we discuss four archetypes.  %Note that the objectives of the designer may be dependent on whether the payoff to an individual for being classified as a $d_i=1$ is $r>0$ (a reward) or $r<0$ (a penalty).  We'll return to this point, but first we define four archetypes.  
The first of these (\textit{accuracy-maximization}) is a standard baseline in statistical decision theory.  The second (\textit{compliance-maximization}) is a standard baseline in many \textit{implementation} problems (such as reducing bad behaviors like fraud and/or promoting good behaviors like physical exercise).  The third and fourth archetypes are less standard, so we begin with a quick definition of accuracy and compliance maximization.

 \subsubsection{Accuracy-Maximization} 
 
 An accuracy-maximizing designer simply seeks to maximize the predictive performance of the algorithm in the sense of maximizing the $\Pr[d_i=\beta_i]$.  This is the probability that the algorithm ``gets it right'' for a randomly drawn individual.   The following defines accuracy-maximizing designers in terms of the parameters for \eqref{dUtility}.
 \begin{definition}
 \label{Def:AccuracyMaximizing}
     The designer $D$ is \textbf{accuracy-maximizing} if $\eta = \{A_1,A_0,B_1,B_0\}$ satisfies the following:
 \[
 A_1=B_1=1 \text{ and } A_0=B_0=0.
 \]
 \end{definition}
As we will return to, note that an accuracy-maximizing designer is indifferent about the individuals' choices of behavior --- such a designer only cares about minimizing the rate of errors produced by the algorithm.

\subsubsection{Compliance-Maximization} 

A compliance-maximizing designer simply seeks to maximize the proportion of individuals who choose $\beta_i=1$.  Such a designer is insensitive to the accuracy of the algorithm's decisions: such a designer's preferences are equivalent to ``the ends justify the means.''  The following defines compliance-maximizing designers in terms of the parameters for \eqref{dUtility}.
 \begin{definition}
 \label{Def:ComplianceMaximizing}
     The designer $D$ is \textbf{compliance-maximizing} if $\eta = \{A_1,A_0,B_1,B_0\}$ satisfies the following:
  \[
  A_1=A_0=1 \text{ and } B_1=B_0=0.
  \]
\end{definition}
As noted earlier, examples of compliance-maximizing incentives would include a designer crafting a ticketing algorithm in order to maximize safe driving, or a fraud detection algorithm designed to minimize fraud.
 
\subsubsection{Moral Hazard}  

The third archetype is a combination of the first two.  It represents a designer who faces no risk (in the sense that his or her payoff is independent of $\beta_i$) if individual $i$ is assigned the decision $d_i=0$, but $D$'s payoff is sensitive to $\beta_i$ if the algorithm assigns $i$ the decision $d_i=1$.   These preferences are very similar to the preferences assumed in models of \textit{moral hazard}.  Such models are widely used, including in models of statistical discrimination (\textit{e.g.}, \cite{CoateLoury93}, \cite{PattyPenn21}).  Formally, a designer facing moral hazard receives 1 from hiring a qualified person, 0 from hiring an unqualified person, and $w\in(0, 1)$ for not hiring ($w$ represents the loss from paying a wage to an unqualified worker).  The following defines a designer facing moral hazard in terms of the parameters for \eqref{dUtility}.
 \begin{definition}
 \label{Def:MoralHazard}
     The designer $D$ faces \textbf{moral hazard} if $\eta = \{A_1,A_0,B_1,B_0\}$ satisfies the following:
\[
A_1=1,\textit{ and } B_1=A_0 =w\in(0, 1), \text{ and }B_0=0.
\]
\end{definition}

\subsubsection{Predatory}  

The final archetype is one of a somewhat pathological decision-maker.  Specifically, $D$ has  \textit{predatory} preferences if $D$'s most-preferred outcome is to \textit{not} give the reward ($d_i=0$) to an individual $i$ who did \textit{not} comply ($\beta_i=0$).  This is ``pathological'' in this setting because, if we conceive of compliance as potentially a social good --- which we will shortly ---  the designer's ordinal preferences about behavior ($\beta_i$) are the opposite of the individuals' common preference with respect to others' behaviors. The following defines a designer facing moral hazard in terms of the parameters for \eqref{dUtility}.
 \begin{definition}
 \label{Def:Predatory}
     The designer $D$ is \textbf{predatory} if $\eta = \{A_1,A_0,B_1,B_0\}$ satisfies the following:
\[
B_1=1 \text{ and } A_1=A_0=B_0=0.
\]
\end{definition}
We refer to a designer with this form of preferences as ``predatory" because it captures the incentives of a designer who benefits from inducing noncompliance. There are several ways this can manifest in practice, including predatory towing,  the issuing of punitive interest rates/fees for late payments, or requiring that a loan be over-secured. 

The four types of designers described above are mutually exclusive though clearly not exhaustive. The  preferences described in Equation \ref{dUtility} can capture a rich array of motivations for the algorithm designer, some of which we will return to later.  We now consider how individuals' behavioral choices will respond to any classifier-reward pair, $(\delta,r)$.  

\section{Responding to the Algorithm: Algorithmic Endogeneity}

In many social science applications, some or all individuals may be aware of the details of the algorithm by which they are classified.  With this awareness, individuals might alter their behaviors in anticipation of being classified by an algorithm. The probability that $D$ classifies an individual sending signal $s_i$ as $d_i=s_i$ is $\delta_{s_i}$. Accordingly, conditional on the algorithm, $\delta$, and the individual cost, $\gamma_i$, if $i \in N$ chooses $\beta_i = 1$, then $i$ receives an expected payoff equal to 
\begin{eqnarray}
EU_i(\beta_i=1\mid \gamma_i, \delta, \phi) & \equiv & E_{d_i}[u_i(\beta_i=1, \cdot \mid \gamma_i, r, t), \nonumber \\
& = & r \cdot (\phi \cdot \delta_1+(1-\phi)(1-\delta_0)) -\gamma_i. \nonumber
\end{eqnarray}
If the individual chooses $\beta_i=0$, then $i$ receives an expected payoff equal to
\begin{eqnarray}
EU_i(\beta_i=0\mid \gamma_i, \delta, \phi) & \equiv & E_{d_i}[u_i(\beta_i=1, \cdot \mid \gamma_i, r, t), \nonumber \\
& = & r \cdot (\phi \cdot (1-\delta_0)+(1-\phi)\delta_1).\nonumber
\end{eqnarray}
Consequently, any individual $i \in N$ will choose $\beta_i=1$ if 
\begin{eqnarray}
EU_i(\beta_i=1\mid \gamma_i, \delta, \phi) & \geq & EU_i(\beta_i=0\mid \gamma_i, \delta, \phi), \nonumber \\
\gamma_i & \leq & r \cdot \left(\delta_1+\delta_0-1\right)(2\phi-1). \label{Eq:GammaEq}
\end{eqnarray}  
The left side of Equation \ref{Eq:GammaEq} is $i$'s direct cost of choosing $\beta_i=1$, and the right side represents the relative benefit to the individual of choosing $\beta_i =1$ versus $\beta_i =0$ in terms of the reward, $r$, the algorithm, $\delta=(\delta_1,\delta_0)$, and the accuracy of the signal, $\phi\in (\frac{1}{2},1]$.  
%\paragraph{Expected Responsiveness of the Algorithm.}  
The right hand side of Equation \ref{Eq:GammaEq} will be central to much of our analysis, so we use it as the basis for defining the \textit{expected responsiveness} of any algorithm $\delta$.
\begin{definition}
\label{Def:ExpectedResponsiveness}
For any $\phi \in (1/2,1]$, the \textbf{expected responsiveness} of any algorithm $\delta$ is defined by the following:
\[
\rho(\delta,\phi) \equiv  \left(\delta_1+\delta_0-1\right)(2\phi-1).
\]      
\end{definition}

In words, the expected responsiveness of an algorithm measures the change in the likelihood that $i$ will be classified as $d_i=1$ by choosing $\beta_i=1$ as opposed to $\beta_i=0$, conditional on the algorithm $\delta$ and the accuracy of the signal, $\phi$. When $\delta_1+\delta_0>1$, individuals are \textit{more} likely to receive reward $r$ conditional on sending a signal of $s_i=1$ than conditional on $s_i=0$; when $\delta_1+\delta_0<1$, individuals are \textit{less} likely.  With this, we can divide algorithms into three categories. 

\begin{definition}
An algorithm $\delta$ is
\begin{itemize}
    \item \textbf{positively responsive} if $\rho(\delta,\phi)>0$,
    \item \textbf{negatively responsive} if $\rho(\delta,\phi)<0$, and 
    \item \textbf{null} (or, a \textbf{null classifier}) if $\rho(\delta,\phi)=0$. 
\end{itemize}  
\end{definition}

The reason for this language is two-fold.  The decisions awarded by a responsive algorithm are correlated with the \textit{signal} received by the algorithm about the individual's choice of behavior.\footnote{Because we have assumed that $\phi>1/2$, this implies that a positively (negatively) responsive algorithm's decisions are positively (negatively) responsive to the behavior chosen by each individual, $\beta_i$.}  This implies that the proportion of individuals who choose to comply ($\beta_i=1$) will be positively  correlated with a positive reward, $r$, if the algorithm is positively responsive. Similarly, negatively responsive algorithms are negatively correlated with the signal.

The third category of algorithms --- null classifiers --- are exactly those algorithms in which the algorithm's decisions are uncorrelated with the signal received by the algorithm and, more importantly, with the behavior chosen by the individual.  As we shall see, when the algorithm is null, individuals choose $\beta_i=1$ if and only if $\gamma_i\leq 0$.  Furthermore, this is the only class of algorithms with this property if $r\neq 0$. (As mentioned above, when $r=0$, \textit{all} algorithms have this property.)

%\paragraph{Equilibrium Prevalence.}  

An immediate implication of Equation \ref{Eq:GammaEq} is that, conditional on $D$'s choice of classifier $\delta=(\delta_1, \delta_0)$ and accuracy, $\phi\in (\frac{1}{2},1]$, the equilibrium fraction of individuals investing in $\beta=1$, or \textbf{equilibrium prevalence}, is  
\begin{equation}
\label{Eq:Equilibrium}
\pi_F(\delta,\phi,r)\equiv F(r\cdot \rho(\delta,\phi)).
\end{equation}
We will see that equilibrium prevalence is central not only to the designer's incentives in designing the algorithm, it is also central to the individuals' preferences over the reward, $r$.

% \subsection{Interpreting the Reward} 

% In Section \ref{Sec:Objectives} we laid out several different objectives of an algorithm designer, and noted that some of those objectives require knowledge of the sign of the reward, $r$, in order to make sense. Suppose, for example, that the designer aims to punish individuals who do not comply (individuals classified as $d_i=0$ pay a penalty of $r_p<0$) and reward individuals who do comply (individuals classified as $d_i=1$ receive a reward of $r_w>0$).  Individuals will therefore comply when \small $$r_w(\phi\delta_1+(1-\phi)(1-\delta_0))+r_p(\phi(1-\delta_1)+(1-\phi)\delta_0)-\gamma_i\geq r_w(\phi(1-\delta_0)+(1-\phi)\delta_1)+r_p(\phi \delta_0+(1-\phi)(1-\delta_1)),$$ \normalsize or when $$\gamma_i\leq (r_w-r_p)(\delta_1+\delta_0-1)(2\phi-1).$$ Comparing this to Equation \ref{Eq:GammaEq}, it's clear that the incentive to comply in this case is equivalent to the agent receiving a reward of $r=r_w-r_p>0$ conditional on $d=1$, and nothing otherwise. Consequently, when considering just the behavioral implications of the rewards and penalties to compliance we can collapse $r_w$ and $r_p$ into reward $r$ without loss of generality.

\subsection{The Designer's Incentives: Fundamentals}

Our first result regarding the designer's algorithmic design problem is that the designer can approximate his or her highest possible payoff if the reward, $r$, is sufficiently large.

\begin{proposition}\label{allCells}
As $r\rightarrow\infty$ the designer can attain an expected payoff arbitrarily close to
\[
\max\bigg[A_1, A_0, B_1, B_0\bigg].
\]
\end{proposition}
\begin{proof}
    Proofs of all numbered results are presented in Appendix \ref{Sec:Proofs}.
\end{proof}
Proposition \ref{allCells} implies that when the reward ($r$) is sufficiently large, the designer can design an algorithm that will channel virtually all outcomes (\textit{i.e.}, all behavior/decision pairs $(\beta_i,d_i)$) into any one of the four cells of the confusion matrix.\footnote{If we assumed that the distribution of costs had compact support (\textit{i.e.}, $F(\gamma_L)=0$ and $F(\gamma_H)=1$ for some pair of finite numbers, $(\gamma_L,\gamma_H) \in \mathbf{R}^2$), then we could replace the ``arbitrarily close'' qualifier with ``equal.''}  Proposition \ref{allCells} follows directly from the assumptions we have made about the individuals' preferences, rather than anything we've assumed about the designer's preferences. The proposition shows that the designer's power to influence outcomes is essentially unbounded if the designer can not only design the algorithm, $\delta$, but also choose the level of the reward, $r$.  Accordingly, the proposition is central to our assumption below that the individuals collectively control the level of the reward.  Before reaching that point, however, we continue to analyze the designer's incentives with a fixed finite reward, $r$.

\subsection{The Designer's Optimal Algorithm}  

We are now in a position to characterize $D$'s optimal algorithm $\delta \equiv (\delta_1, \delta_0)$ given that the algorithm will affect the equilibrium prevalence.  The designer's \textit{expected} payoff from any classifier, $\delta$, is:
\small
\begin{equation}
\label{Eq:ObjectiveD}
\begin{array}{l}EU_D(\delta\mid r, F, \phi, \eta) = \\
\mbox{}
\end{array}\begin{array}{l}
\pi_F(\delta,\phi,r)(\phi\cdot (A_1\delta_1+A_0(1-\delta_1))+(1-\phi) (A_0\delta_0+A_1(1-\delta_0)) )\\
+ (1-\pi_F(\delta,\phi,r))(\phi\cdot (B_1\delta_0+B_0(1-\delta_0))+(1-\phi) (B_0\delta_1+B_1(1-\delta_1)) ).
\end{array}
\end{equation}

\normalsize
%\textcolor{blue}{JWP: Should we think about whether we can write this more succinctly?  More substantively, I think that predictive parity is hiding in here.}

With \eqref{Eq:ObjectiveD} in hand, we can prove our first result, which characterizes the optimal algorithm for a designer who only cares about the effect of the algorithm on individual behavior (in other words, a designer seeking to either maximize or minimize compliance).  Such a designer in our setting has a particularly simple optimal algorithm.  As stated in the next proposition, a compliance-maximizing or minimizing designer's optimal algorithm is always degenerate.
\begin{proposition}[Optimal Compliance-Maximizing and Compliance-Minimizing Algorithms]
    \label{Pr:OptimalComplianceAlgorithm}
    If the designer's preferences $\eta$ satisfy $A_1=A_0=\bar{A}\geq 0$ and $B_1=B_0=\bar{B}\geq 0$, then the optimal algorithm depends on $r$ and is defined by the following:
    \begin{equation}
    \label{Eq:OptimalComplianceAlgorithm}
    \delta^*(r,F,\phi,\eta) = \begin{cases}
        (1,1) & \text{ if } \;\; r\cdot (\bar{A}-\bar{B}) > 0, \\
        (0,0) & \text{ if } \;\; r\cdot (\bar{A}-\bar{B}) < 0.
    \end{cases}
    \end{equation}
    and all algorithms are equivalent to the designer if $r=0$ or $\bar{A}=\bar{B}$.
\end{proposition}

Our first general result (in the sense of not depending on the designer's preferences) is that any optimal classifier will be a corner solution in either $\delta_1$, or $\delta_0$, or both.%\footnote{As a side note, Proposition \ref{Pr:OptimalComplianceAlgorithm} is actually proven in the proof of Proposition \ref{Pr:Corner}, which demonstrates the central role of the linearity of the designer's preferences from an analytical standpoint.}

\begin{proposition}[Optimal Algorithms Are Almost Never Interior]
\label{Pr:Corner}
When $r\neq 0$, any optimal classification strategy for $D$ requires either $\delta_1^*\in \{0, 1\}$, or $\delta_0^*\in\{0, 1\}$, or both. %When $\phi=0.5$, this holds \emph{unless}
%\[
%F(0)A_1+(1-F(0))B_0=F(0)A_0+(1-F(0))B_1,
%\]
%in which case any classification strategy is optimal for $D$. 
\end{proposition}

Our next characterization is partial in the sense that it holds for only a subset of all possible preferences for the designer.\footnote{And, to be clear, we have also assumed that the distribution of costs (types) has a log-concave probability density function, $f$.}  The cases we focus on here are those in which the designer has an unambiguous preference regarding the accuracy of the algorithm's decisions conditional on individual behavior.  We refer to these cases as \textit{accuracy aligned} or \textit{accuracy misaligned} designers.  We consider accuracy alignment to be a natural condition, and indeed it is satisfied by each of the four families of designer objectives defined in Section \ref{Sec:Objectives}.

\begin{definition}
\label{Def:AccuracyAlignment}
The designer, $D$, is 
\begin{itemize}
    \item \textbf{accuracy aligned} if $A_1\geq A_0$ and $B_1\geq B_0$, and \textbf{strongly accuracy aligned} if at least one inequality is strict, and
    \item \textbf{accuracy misaligned} if $A_1\leq A_0$ and $B_1\leq B_0$, and \textbf{strongly accuracy misaligned} if at least one inequality is strict.
\end{itemize}
\end{definition}

If the designer is both accuracy aligned and misaligned then the designer must be either compliance-maximizing or minimizing.  The notions of accuracy alignment or misalignment are useful to us, because they cleanly identify a key aspect of the algorithm designer's optimal algorithm.  As stated in the next proposition, any designer whose preferences are strongly accuracy aligned or misaligned should design an algorithm that is degenerate (\textit{i.e.}, uses a pure strategy) with respect to at least one of the two possible signals, $s_i \in \{0,1\}$.  This is stated formally in the following proposition.

\begin{proposition}
\label{bigConcavity}
If $D$ is strongly accuracy aligned or misaligned,  then $D$'s payoff is strictly quasiconcave in $\delta_j$ and strictly quasiconvex in $\delta_{1-j}$, for some $j \in \{0,1\}$. 
\end{proposition}
From a technical perspective, Proposition \ref{bigConcavity} is  useful because it reduces the search for the designer's optimal algorithm to a one-dimensional optimization problem for any fixed reward, $r$. Moreover, the sign of $r$ and the accuracy alignment or misalignment of $D$'s preferences pin down the quasiconvexity/concavity properties of $\delta_1$ and $\delta_0$ (these properties are characterized in the proof of Proposition \ref{bigConcavity}).

From a substantive standpoint, Proposition \ref{bigConcavity} is informative: an accuracy-aligned algorithm designer has an instrumental incentive to remove as much ``noise'' in the awarding of decisions as possible, and this leads to a very robust conclusion that at most one of the two signals will leave any residual uncertainty (\textit{i.e.}, ``deliberately induced noise'') about the decision that will be subsequently rendered by the algorithm.

\subsection{A Motivating Example: Maximizing Accuracy is Not Neutral \label{accNeutral}}

We'll illustrate some of the takeaways of our model of optimal classification by considering a designer who solely seeks to maximize accuracy: to reward individuals that comply, and to penalize individuals that don't.  For the purposes of this example we set the reward at $r=10$, and the signal accuracy at $\phi=\frac{3}{4}$.  In this case the designer's objectives can be described by $A_1=B_1=1$ and $A_0=B_0=0$.  For comparability, we depict this payoff function with the confusion matrix displayed in Table \ref{Tab:AccuracyIsNotNeutralExample}.  

\begin{table}[hbtp]
    \centering
\begin{tabular}{|c||c|c|} \hline
&\multicolumn{2}{c|}{Decision} \\ \hline
Compliance&$d_i=1$&$d_i=0$ \\ \hline \hline

\multirow{2}{*}{$\beta_i=1$}&$A_1=1$&$A_0=0$ \\ 
& (True Positive)&(False Negative) \\ \hline
\multirow{2}{*}{$\beta_i=0$}&$B_0=0$&$B_1=1$\\ 
& (False Positive) & (True Negative) \\ \hline
 \end{tabular}
 \caption{Pure Accuracy-Maximizing Payoffs}
    \label{Tab:AccuracyIsNotNeutralExample}
\end{table}
Consider first the designer's optimal classifier when costs to compliance, $\gamma$, are distributed according to the $\mathrm{Normal}(0,1)$ distribution.  In this case, a null classifier (\textit{i.e.}, any $\delta = (\delta_1, 1-\delta_1)$) will yield an equilibrium prevalence equal to 
\[
\pi_F(\delta,\phi,r)=F(0)=\frac{1}{2}.
\]
Accordingly, any null classifier in this setting will yield $D$ an expected payoff equal to $\frac{1}{2}$. 

If, on the other hand, $D$ simply followed the signal, using the degenerate positively responsive algorithm $\delta=(1,1)$, then $D$ will receive an expected payoff of $\pi_F(\delta,\phi,r) = \phi = \frac{3}{4}$. Thus, no null classifier is optimal for $D$ in this case, as $D$ can do better with  $\delta=(1,1)$. However, note the equilibrium prevalence induced by the degenerate positively responsive algorithm: 
\[
\pi_F(\delta,\phi,r)=F(r\cdot (2\phi-1))=F(5) \approx 1.
\]
In other words, almost every individual is complying but the algorithm is penalizing $1-\phi=25\%$ of these complying individuals. %\footnote{Similarly, this algorithm is rewarding $1-\phi$ of the (nearly negligible) set of individuals who choose not to comply.}  
From an ex post perspective --- or, in other words, holding the observed prevalence constant --- $D$ would strictly benefit from using the classifier $\delta=(1, 0)$ that rewards \textit{all} individuals, regardless of the signal.

Clearly this approach will not benefit $D$ in equilibrium (\textit{i.e.}, taking into the fact that algorithm will determine equilibrium prevalence), because this alternative classifier is a null classifier.  As discussed above, any such algorithm would induce all individuals with positive costs to not comply, so that equilibrium prevalence would drop to $\pi_F(\delta,\phi,r)=\frac{1}{2}$, and $D$ would receive an equilibrium payoff equal to $\frac{1}{2}$.

%\paragraph{The Optimal Algorithm in this Setting.} 
Given our choice of $r=10$ and $\phi=0.75$, $D$'s optimal classifier is 
\[
\delta^*=(1,0.37). 
\]
This algorithm rewards all individuals who send a signal of $s_i=1$,  and rewards about $63\%$ of all individuals who send a signal of 0.  This classifier incentivizes fewer individuals to comply than if the designer simply followed the signal (97\% versus nearly 100\%), but it accurately classifies almost 90\% of the population.

\begin{figure}[hbtp]
\centering
\framebox{
\epsfig{file=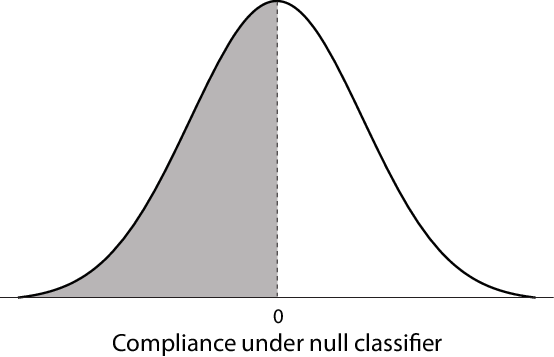, scale=.5}\hspace{.5in} \epsfig{file=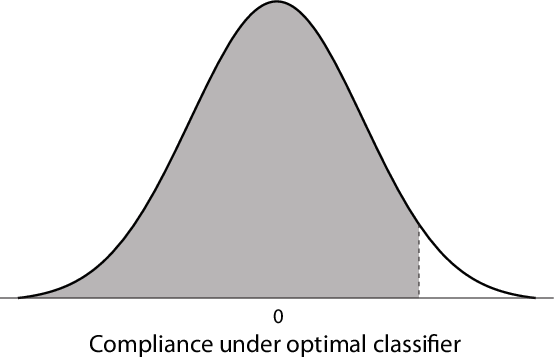, scale=.5}}
\caption{Optimal Accuracy-Motivated Algorithm Induces Compliance}
\label{Fig:AccuracyIsNotNeutral}
\end{figure}

Figure \ref{Fig:AccuracyIsNotNeutral} displays the distribution of behaviors under a null classifer (left panel) and $D$'s optimal algorithm (right panel), with the mass of compliers shaded gray.

\begin{figure}[hbtp]
\centering
\framebox{
\epsfig{file=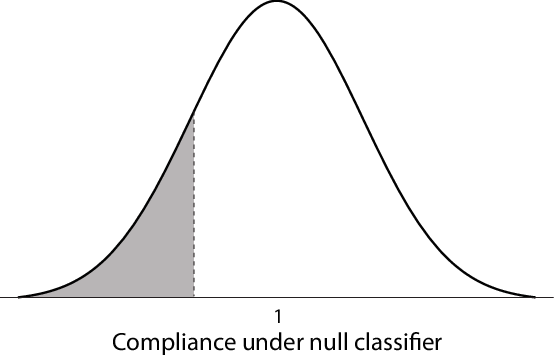, scale=.5}\hspace{.5in} \epsfig{file=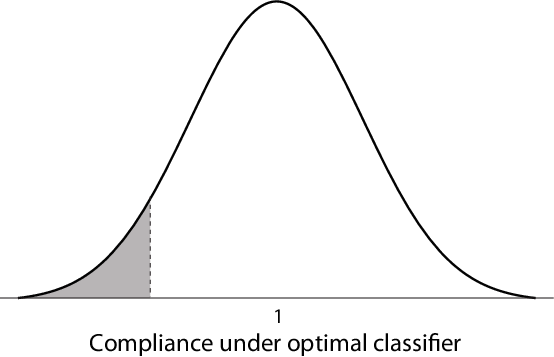, scale=.5}
}
\caption{Optimal Accuracy-Motivated Algorithm Induces Non-Compliance}
\label{Fig:AccuracyIsNotNeutral2}
\end{figure}

%\paragraph{Implications.}  
The point to take from this example is that in order to \textit{most accurately classify} compliance, the designer's optimal classifier \textit{induces} compliance.  This may be good if we assume that aggregate compliance is socially desirable. However, an accuracy-motivated designer need not induce this kind of desirable outcome.
%\paragraph{Sensitivity to the Cost Distribution.}  
To see this, consider the same setting as above with one change: the mean of the distribution of the cost of compliance has shifted rightward, such that the costs of compliance are distributed according to the $\mathrm{Normal}(1, 1)$ distribution.  In this case, the optimal classifier for an accuracy-motivated designer is $\delta_1=0$ and $\delta_0=0.92$: all individuals sending a signal of 1 are penalized, and 92\% of the individuals sending a signal of 0 are penalized.  This classifier \textit{disincentivizes compliance}, with fewer individuals complying than under a null classifier (8\% versus 16\%).  It is highly accurate however, and again correctly classifies almost 90\% of individuals.  Mirroring Figure \ref{Fig:AccuracyIsNotNeutral}, Figure \ref{Fig:AccuracyIsNotNeutral2} illustrates how the same accuracy-maximizing designer will find it optimal in this case to effectively incentivize \textit{non-}compliance.

\subsection{Why is Accuracy Not Neutral?}

The examples above demonstrate that a designer who is \textit{solely interested in accuracy} may have induced preferences over behavior.  The reason for this stems from the effect of the algorithm on individual behavior. We often think of accuracy as being defined with respect to a \textit{fixed} target.  However, the general definition of accuracy is agnostic about the nature of the target and, in this setting (and all settings with algorithmic endogeneity, in which individuals care about how they are classified), the location of ``the target'' for an accuracy maximizing designer may be a function of the algorithm chosen by the designer.  When this is the case (as it is in our setting), an accuracy-maximizing designer has an incentive to design an algorithm that effectively makes the target ``easier to hit.''  This second-order incentive leads  an accuracy-maximizing designer to choose an algorithm that can compensate for the inherent noisiness of the signal by inducing individuals to behave in the same way.
This is at odds with designer incentives in a setting in which the prevalence is treated as exogenous to the details of the algorithm itself.  % \textcolor{red}{BOOBOO. I don't really get this. The designer would choose (1, 1) in the N[0, 1] case here, because $\pi_F=\frac{1}{2}$. So with exogenous compliance, the designer is identical to the compliance-minded designer with endogenous compliance. I'm missing something!}

\section{Democratic Rewards}

So far, we have shown that the designer of a classification algorithm can exert considerable control over societal outcomes.  As the previous sections demonstrated, this control is not only with respect to how individuals are classified as deserving of a reward or penalty, but also with respect to the behavior individuals optimally engage in.  A designer seeking to maximize ticketing revenue will induce very different aggregate societal behavior than a designer seeking to maximize public safety, even when considering two populations that have the same underlying costs to compliance.  Moreover, Proposition \ref{allCells} demonstrates that control of the size of the reward, $r$, would grant the designer the essentially unfettered 
ability to achieve his or her preferred outcome for every individual.  Motivated by recent democratic reforms to change the stakes of algorithmic classification, we now explore the outcomes and rewards each individual $i \in N$ would most prefer for a given algorithm, $\delta$.

We consider a setting where rewards $r$ are chosen by the community via a majority vote.  We assume, in line with the seminal model of a rational size of government by \cite{MeltzerRichard81}, that the rewards must be financed equally (if $r>0$) by, or the fines must be distributed equally (if $r<0$) back to, all individuals.  Substantively, this is a budget balance requirement.  However, the main reason to make this assumption is technical: if there is no budget constraint, then all individuals would weakly prefer higher levels of rewards.\footnote{This issue is discussed in \cite{PattyEPG08} in the related context of how legislators might create incentives to maintain party unity.}

% Voters receive $r$ if classified as $d_i=1$, and this reward is financed by a tax borne by each voter of $r\overline{d}$, where $\overline{d}=E[d_i]$, or: $$\overline{d}=F(r\cdot\rho(\delta, \phi))(\phi \delta_1+(1-\phi)(1-\delta_0))+(1-F(r\cdot\rho(\delta, \phi)))(\phi(1-\delta_0)+(1-\phi)\delta_1).$$ If we want to alternatively conceive of the voters as determining a system of rewards and penalties (as opposed to rewards financed through taxation), we can think of the reward of being classified as a $d_1=1$ as $r_w=r(1-\overline{d})$, and the penalty to being classified as a $d_i=0$ as $-\overline{d}$.  

%\paragraph{Why Would Individuals Want to Subsidize Rewards?}  
We amend the payoff function for individuals given in Equation \ref{Eq:BaselineVoterPayoff} to accommodate our balanced budget constraint by requiring that each individual pay a tax equal to the average reward that is awarded to individuals, $r\cdot \overline{d}$. %With 
%\begin{center}$\begin{array}{ccl}
%E[d]&=&F(r\cdot \rho(\delta, \phi)) (\phi \delta_1+(1-\phi)(1-\delta_0)) +(1-F(r\cdot \rho(\delta, \phi)))(\phi(1-\delta_0)+(1-\phi)\delta_1)\\
%&=&\delta_1 F(r\cdot \rho(\delta, \phi))+(1-\delta_0)(1-F(r\cdot \rho(\delta, \phi))).\end{array}$\end{center}  
This tax is equivalent to assuming that people receive reward $r\cdot(1-\overline{d})$ if $d_i=1$, and pay penalty $-r\cdot \overline{d}$ if $d_i=0$.

Additionally, we also add a term to the payoff function that allows individuals to have a taste aggregate behavior.  Letting $\pi$ denote the prevalence of compliance in the population, we capture this term by $t\cdot \pi$, for $t\geq 0$.  Neither this term nor the tax $r\cdot \overline{d}$ affect an individual's decision to comply, and so they do not change any of our results thus far.  However, these terms do affect preferences over optimal rewards and penalties. 

Incorporating a taste for aggregate compliance into individuals' payoffs allows us to consider that individuals, as members of a common community, may share preferences over aggregate behavior.  The marginal value of the (positive) externality generated by others' choices to comply ($\beta_j=1$) is represented by $t \geq 0$. As $t$ increases, all individuals value aggregate compliance more, which can be conceived as an increased negative externality of non-compliance.  Residents of a dense urban community may, for example, value safe driving in the aggregate more than residents of a rural community.  We will see that as $t$ increases, all individuals become more supportive of subsidizing compliance.  That said, individual tastes for this subsidy also depend on their private costs to compliance.  Consequently, even when $t=0$ every individual will prefer a system of positive rewards and fines. Incorporating these terms into individuals' payoffs, individual preferences are now described by \begin{equation}\label{newPrefs}
r\cdot (d_i-\overline{d})-\beta_i\cdot \gamma_i+t\cdot \pi.
\end{equation}

%Given the imposition of budget balance and the assumption that individuals have linear preferences over the reward, it is useful to consider why any individual might have a strict preference for subsidizing a reward for other individuals.  

%The ex-post payoffs of voters are: $$U_V(r)=t  F(r\cdot\rho(\delta, \phi))+r d_i-r \big(F(r\cdot\rho(\delta, \phi))(\phi \delta_1+(1-\phi)(1-\delta_0))+(1-F(r\cdot\rho(\delta, \phi)))(\phi(1-\delta_0)+(1-\phi)\delta_1)\big).$$  The term $d_i$ is the decision that the algorithm assigns voter $i$, and is consequently a probabilistic function of behavior $\beta_i$. 

We assume that the structure of the problem is common knowledge to all individuals (including the designer).\footnote{This doesn't preclude the possibility that individuals have privately observed types, but our analysis also clarifies that, because we require the algorithm designer to use the algorithm to render individual decisions, it is not important whether the designer is aware of any given individuals' types, because the algorithm is not allowed to condition upon this information.
% The (or at least most) individuals would prefer that the algorithm _could_ condition upon this information.
}
We analyze equilibrium behavior, and the starting point of this analysis is to consider how each individual $i$ should calculate his or her most-preferred reward level.  Each individual will realize that he or she will ultimately choose either to comply or not.  Conditional on each of these possible choices, the distribution of types, $F$, the algorithm, $\delta$, and individual $i$'s type, $\gamma_i$, $i$ calculates his or her most-preferred reward in each of the two cases.  This yields the following conditional expected payoff function for any given individual $i \in N$:\footnote{Equation \eqref{twoPeak} is derived in Appendix \ref{Sec:TechnicalStuff} (Equations \eqref{rInvest} and \eqref{rNoInvest}).}
\begin{equation}
EU_V(r|\gamma_i, \delta, t, F, \phi)=\begin{cases} 
-\gamma_i+r\cdot\rho(\delta, \phi)(1-F(r\cdot\rho(\delta, \phi)))+t\cdot F(r\cdot\rho(\delta, \phi)) \,\,\,\text{ if }\gamma_i\leq r\cdot \rho(\delta, \phi),\\
-r\cdot\rho(\delta, \phi) F(r\cdot\rho(\delta, \phi)) +t\cdot F(r\cdot\rho(\delta, \phi)) \,\,\,\text{ otherwise.}
\end{cases}
\label{twoPeak}\end{equation}

The following proposition establishes that each individual's maximization problem is well-defined.
\begin{proposition} If $\delta$ is not null, then conditional on behavior $\beta_i$, voter payoffs are strictly quasi-concave in rewards, $r$, and maximized at an interior $r$. If $\delta$ is null, each voter $i$ is indifferent between all reward levels.
\label{intR}
\end{proposition}
The following corollary presents the two potentially optimal rewards for any individual $i$ (one conditional on $i$ subsequently choosing to comply, and the other conditional on $i$ subsequently choosing to not comply).  

\begin{corollary}\label{eqRCorr}
The optimal $r_1^*$ and $r_0^*$ (rewards for individuals choosing to comply and not comply respectively) are of the form: 
\begin{equation}
    \label{Eq:HighRewardSupport}
r_{\beta_i}^*(\delta\mid t, F, \phi)=\frac{k_j^*(t,F)}{\rho(\delta,\phi)} %= \frac{k_j^*(t,F)}{(\delta_1+\delta_0-1)(2\phi-1)},
\;\;\; \forall \beta_i \in \{0,1\},
\end{equation}
with the values $k_0^*(t,F)$ and $k_1^*(t,F)$ defined implicitly as follows:

\begin{eqnarray*}
k_0^*(t,F) & = & t-\frac{F(k_0^*(t,F))}{f(k_0^*(t,F))},\\
\\ \vspace{.1in}
k_1^*(t,F) & = & t+\frac{1-F(k_1^*(t,F)) }{f(k_1^*(t,F))}.
\end{eqnarray*}
\end{corollary}

\ \\Corollary \ref{eqRCorr} shows that, for any given $\delta$, $t$, $F$, and $\phi$, there are only two possible ideal rewards --- $r_0^*$ or $r_1^*$ --- for any given individual $i$.  Furthermore, \textit{these two possible ideal reward levels are identical across all individuals}.  This is the combined result of the assumption of budget balance and the assumption that all individuals have a common marginal preference for compliance by others ($t$ is common to all).  That said, what is especially surprising about this is that individuals are not homogeneous --- they each know their own types.

Whenever the context is clear, we will omit the arguments of $k_0^*$, $k_1^*$, $r_0^*$, and $r_1^*$.  With the optimal $r_i^*$ derived, it can be shown that a voter with costs $\gamma_i$ receives an expected payoff  from $r=r^*_1$ that is at least as great as from $r=r_0^*$ if and only if
\[
\gamma_i\leq k_0\cdot F(k_0)+k_1\cdot(1-F(k_1))+t\cdot(F(k_1)-F(k_0)).
\]
Because the optimal rewards, $r_1^*$ and $r_0^*$, are characterized by the terms $k_1$ and $k_0$, we define the following term $k_i^*$, which we refer to as individual $i$'s ``optimal $k$.''
\begin{equation}
\label{kStar} 
k_i^*=\begin{cases} k_1\,\,\,\text{ if }\gamma_i\leq k_0\cdot F(k_0)+k_1\cdot(1-F(k_1))+t\cdot(F(k_1)-F(k_0)),\\
k_0 \,\,\,\text{ otherwise.}
\end{cases}
\end{equation}

The value $k_i^*$ can be interpreted essentially as the optimal level of compliance, given $i$'s type, $\gamma_i$, because $F(k_i^*)$ represents $i$'s optimal \textit{equilibrium prevalence}.  Individuals with $k_i^*=k_1$ prefer higher prevalence (in equilibrium, \textit{i.e.}, after taking transfers and the distribution of others' costs into account) than individuals with $k_i^*=k_0$.  We shall see that, in all equilibria with non-null algorithms, some individuals will ``vote for'' high compliance but ultimately not comply, or vice-versa.  We will return to this, but the point to note is that this seeming preference reversal will be solely a function of the individual in question being on the ``losing side'' of the majority vote over the ultimate reward.  

For any given pair $(k_0^*,k_1^*)$, Equation \eqref{kStar} defines a cut-point that divides individuals (in terms of their types) into ``low cost'' and ``high cost'' individuals --- individuals with low enough costs will support the higher reward level, $r_1^*$, and individuals with high costs will support the lower reward level, $r_0^*$.  Equation \eqref{kStar} also demonstrates, as claimed earlier, that support for the higher reward increases in the marginal value of the externality, $t$.  This is stated formally in the following proposition.

\begin{proposition}
\label{symmetricDist}
For any $t$, $F$, and $\phi$, and any voter $i \in N$,
\[
\bigg[ \; k_i^*(t,F)=k_1^*(t,F) \; \text{ \emph{and} } \; t'>t \; \bigg] \Rightarrow \;\; k_i^*(t',F)=k_1^*(t',F).
\]
\end{proposition}

%BOOBOO I removed this because I don't know that it's true or see why we need it right now.
%A similar ordering can be established with respect to the first order stochastic dominance partial order on cost distributions, $\mathcal{F}$.  Write $F \succ F'$ if the distribution $F$ first order stochastically dominates $F'$.  The next proposition establishes that lowering the costs of compliance in this sense will not induce any voter to support lower equilibrium compliance.
%\begin{proposition}
%\label{Pr:FOSDOrderingOfKStar}
%For any $t$, $F$, and $\phi$, and any voter $i \in N$,
%\[
%\bigg[ \; k_i^*(t,F)=k_1^*(t,F) \; \text{ \emph{and} } \; F \succ F' \; \bigg] \Rightarrow \;\; k_i^*(t,F')=k_1^*(t,F').
%\]
%\end{proposition}

With the comparative statics of individual incentives established, we now turn to the question of how rewards will be chosen democratically for any given classifier $\delta$.

\subsection{A Median Voter Theorem}

Our first result is that there is always a Condorcet winner among rewards for any distribution of types $F$, precision $\phi$, and algorithm $\delta$.  Specifically, recalling that the individuals are indexed by the unit interval, $N=[0,1]$, and ordered by their individual costs, $i\leq j \Leftrightarrow \gamma_i \leq \gamma_j$, individual $i=0.5$'s cost of complying, $\gamma_{0.5}$, is equal to the median of the distribution of individual costs.  We denote this individual by $\mu$, and the next proposition states that, for any classifier, $\delta$, individual $\mu$'s ideal reward is a Condorcet winner among all possible reward levels.
\begin{proposition}
\label{Pr:MedianVoterTheorem}
For any classifier, $\delta$, marginal value of compliance, $t$, distribution $F$, and precision $\phi$, the reward 
\[
r^*(\delta\mid t, F, \phi) = \frac{k_\mu^*(t,F)}{\rho(\delta, \phi)},
\]
is a Condorcet winner: it is preferred by a majority of individuals to \textit{any} other reward, $r \in \mathbf{R}$.
\end{proposition}
The proof of Proposition \ref{Pr:MedianVoterTheorem} (in Appendix \ref{Sec:Proofs}) is straightforward, because Corollary \ref{eqRCorr} ensures that there are only two ideal rewards for any non-null algorithm, and all voters are indifferent regarding the reward level for any null classifier. 

%BOOBOO

We can now describe some basic properties of outcomes when rewards are set democratically.  Referring to the fraction of individuals complying when rewards are democratically set as \textbf{democratic compliance}, the first result is a corollary of Proposition \ref{Pr:MedianVoterTheorem}, but has far-reaching implications.  Specifically, for any given $F$ and $\phi$, democratic compliance is insensitive to the design of any non-null algorithm.  In other words, the only aspect of the algorithm that can affect democratic compliance in equilibrium is whether the algorithm is null or not.  

\begin{corollary}\label{optimalPrevalence} For any non-null classifier, democratic equilibrium compliance is equal to $F(k_\mu^*)$. For any null classifier, equilibrium compliance (regardless of how $r$ is set) is equal to $F(0)$.
\end{corollary}

The next result strengthens Corollary \ref{optimalPrevalence} --- it does not follow immediately from the corollary because it is possible that voters could have strict preferences over different algorithms because of the expected transfers that will occur in equilibrium.  Proposition \ref{Pr:VoterIndifferenceNonNull} clarifies that the invariance of compliance with any non-null algorithm translates seamlessly into indifference over all non-null algorithms.  Furthermore, all voters are indifferent over all null algorithms.  
\begin{proposition}
\label{Pr:VoterIndifferenceNonNull}
When rewards are chosen democratically, every voter is indifferent between all non-null classifiers. Regardless of how rewards are chosen, every voter is indifferent between all null classifiers.  
\end{proposition}
Note that voters will, in general, have a strict preference between non-null and null algorithms.  The main impact of Proposition \ref{Pr:VoterIndifferenceNonNull} for our purposes is that it clarifies that the voters' induced preferences over designers will depend entirely on whether the designer will result in a null or non-null equilibrium.  

%\paragraph{When does the Median Voter Want to Comply?}  

Our final result concerning voter preferences over rewards gives us some insight into when democratic rewards are comparatively high (set at $r_1$, with the median complying) or comparatively low (set at $r_0$, with the median not complying).  If we suppose that the cost distribution $F$ is symmetric about its mean, then the median voter will prefer the higher reward --- and correspondingly, will comply in equilibrium --- if his or her cost (\textit{i.e.}, $\gamma_\mu$) is less than or equal to the marginal value of the externality, $t$. 
 The following proposition states this formally.
\begin{proposition}
\label{Pr:MedianCutpoint} 
If $f$ is log-concave and symmetric about its mean, then the median voter ($i=\mu$) receives a higher payoff at $r_1^*$ than $r_0^*$ if and only if $\gamma_\mu\leq t$.
\end{proposition}
Proposition \ref{Pr:MedianCutpoint} relies on the supposition that $F$ is symmetric only in order to make the statement as clean as possible --- all voters' preferences are continuous with respect to the distribution of the voters' types, so deviating from symmetry will not radically alter the proposition's conclusion.  We now proceed to put the algorithm designer's and voters' problems together and consider a general equilibrium model of algorithm design and democratic reward choices.

%\begin{remark}
 %   Proposition \ref{Pr:MedianCutpoint} mirrors the central feature of equilibrium voting in \cite{MeltzerRichard81}.  Any voter $i$ (including the median voter) in their model will vote for a regressive tax when $i$'s income is higher than the mean income.  Similarly, in our model, any voter $i$ (including the median voter, $i =\mu$) will vote for the reward that is optimal conditional on  \textit{not} complying if $i$'s cost of complying is greater than the mean cost of compliance.
%\end{remark}

\subsection{Democratic Algorithmic Equilibrium}

Returning to the designer's problem of designing an optimal classifier, Corollaries \ref{eqRCorr} and \ref{optimalPrevalence} simplify our problem considerably, because when voters have a say in the system of rewards and punishments a classifier metes out, the fraction of individuals choosing behavior $\beta_i=1$ is either $F(0)$ (a consequence of a null classifier, or a reward of $r=0$, or both), or $F(k_\mu^*)$ in the event that the classifier is non-null. We'll simplify things by assuming throughout that $D$'s preferences are accuracy aligned, or $A_1\geq A_0$ and that $B_1\geq B_0$. As noted earlier, this is a condition that all of our vignettes satisfy and it enables us to pin down the concavity and convexity properties of the designer's objective function via Proposition \ref{bigConcavity}.

We now consider a general equilibrium problem in which, in equilibrium, $r^*$ is chosen to maximize the payoff of the median voter conditional on a choice of algorithm $\delta^*$, and $\delta^*$ maximizes the payoff of the designer conditional on the median's choice of $r^*$.  First, recalling that $\gamma_\mu$ is the median of the individuals' costs of complying, we define our equilibrium concept as follows.

\begin{definition} \label{Def:Equilibrium}
For any $(t,F,\phi,\eta)$, an algorithm-reward pair, $(r^*,\delta^*)$ is an \textbf{equilibrium} if both of the following hold:
\begin{itemize}
\item $r^*(\delta^* \mid t, F, \phi)\in\argmax\limits_{r\in\mathbf{R}} EU_V(r|\gamma_\mu, \delta^*, t, F, \phi)$, and
\item $\delta^*(r^* \mid F,\phi,\eta) \in\argmax\limits_{\delta\in[0, 1]\times[0, 1]} EU_D(\delta|r^*, F, \phi, \eta)$.
\end{itemize}
\end{definition}
In words, the first of the two conditions in Definition \ref{Def:Equilibrium} requires that, conditional on $D$'s choice of algorithm, the reward is equal to the Condorcet winner among all rewards.  The second condition requires that, conditional on the median voter's most-preferred reward, the algorithm designer is choosing an optimal classifier given $D$'s preferences.

\subsubsection{Equilibrium Existence \& Characterization}  We denote the \textbf{equilibrium classifier-reward correspondence}, given $t,F,\phi$, and $D$'s preferences, $\eta \equiv \{A_1,A_0,B_1,B_0\}$, by $\mathcal{E}(t,F,\phi,\eta)$.\label{Pg:EquilibriumCorrespondence}   We begin by considering the existence of a ``null" equilibrium in which $r^*=0$ and/or $\delta^*$ is null.

% We know that in any equilibrium in which the median voter is maximizing his payoff in response to a classifier $\delta$, we have $r^*={{k_\mu^*}\over{(\delta_1+\delta_0-1)(2\phi-1)}}$.  Similarly, we know that when $r^*>0$ then $\delta_1^*\in\{0, 1\}$, and payoffs are strictly quasiconcave in $\delta_0^*$ for any $\delta_1$.  Consequently $\delta^*_0$ is either a corner solution or solves ${\partial\over{\partial \delta_0}}EU_D(\delta\mid r, F, \phi \eta)=0$. (When $r^*<0$ the concavity and convexity properties of $D$'s payoff at $\delta^*_1$ and $\delta^*_0$ are flipped, and so $\delta_0^*\in\{0, 1\}$, etc.).

% \ \\To find the equilibrium classifier, we set up the first order conditions for a general $r$ and $\delta$.  Then we utilize the fact that $r^*={{k_\mu^*}\over{(\delta_1+\delta_0-1)(2h-1)}}$ (so that a fixed fraction of individuals $F(k_\mu^*)$ invest in $\beta_i=1$) and solve for the optimal $\delta^*$. % \textcolor{red}{Note that incorporating the $r^*$ prior to performing $D$'s optimization problem results in a Stackelberg equilibrium, and we don't want this.}

Our first result is that a null equilibrium exists if and only if either (1) the median voter's preferred level of compliance is exactly equal to the level of sincere prevalence ($F(0)$) or (2) sincere prevalence is sufficiently high or low.\footnote{Generically, the median voter's preferred level of compliance will differ from the sincere preference, so the second case is the more important of the two.} This is stated formally in the next proposition.

\begin{proposition}
\label{Pr:NullExistence}
When $k_\mu^*\not=0$, there always exists a null equilibrium when: 
\[
F(0)\not\in \left( {{(B_1-B_0)(1-\phi)}\over{(B_1-B_0)(1-\phi)+(A_1-A_0)\phi}},{{(B_1-B_0)\phi}\over{(A_1-A_0)(1-\phi)+(B_1-B_0)\phi}}\right).
\]
Otherwise, there never exists a null equilibrium.
\end{proposition}

 Proposition \ref{Pr:NullExistence} leads immediately to the following two corollaries:

\begin{corollary}\label{predatoryExist}
If the designer's preferences are of the form $$\eta\in\{\{A_1, 0, 0, 0\}, \{0, A_0, 0, 0\}, \{0, 0, B_1, 0\}, \{0, 0, 0, B_0\}\}$$ then a null equilibrium always exists.
\end{corollary} 

 Corollary \ref{predatoryExist} implies that when the designer only places positive value on at most one cell of the confusion matrix (such as in our vignette describing ``predatory" designer preferences) then there always exists a null equilibrium.

\begin{corollary}\label{highCostExist} If $\phi<1$ then there always exists a null equilibrium when  $F(0)$ is sufficiently low or sufficiently high.
\end{corollary}

 Corollary \ref{highCostExist} is important because it demonstrates that an equilibrium exists for a large class of relevant settings: those in which (virtually) every individual pays some positive cost to compliance. 

Finally, Proposition \ref{Pr:NullExistence} has another implication: null equilibria become less likely to exist as $\phi \to 1$. Turning this around, Proposition \ref{Pr:NullExistence} implies that the equilibrium --- if one exists --- is more likely to involve a positively or negatively responsive algorithm as the algorithm's ``data'' becomes more precise.\footnote{This is related to the point raised by \cite{PattyPenn23PerfectPrediction} regarding the social efficiency of at least a little imprecision/noise in situations of algorithmic endogeneity.}

% \ \\We will denote a null equilibrium by  $(r_\emptyset^*, \delta_\emptyset^*)$, with $r_\emptyset^*=0$. Assuming $A_1-A_0+B_1-B_0\not=0$, $$\delta_\emptyset^*=\begin{cases}
% (1, 0)\text{ if }F(0)> {{B_1-B_0}\over{A_1-A_0+B_1-B_0}},\\
% (0, 1)\text{ if }F(0)< {{B_1-B_0}\over{A_1-A_0+B_1-B_0}},\\
% (\delta, 1-\delta)\text{ for any }\delta\in[0, 1]\text{ otherwise.}
% \end{cases}$$ If $A_1-A_0+B_1-B_0=0$ then, by our assumption that $A_1\geq A_0$ and $B_1\geq B_0$ it must be that $A_1=A_0$ and $B_1=B_0$. In this case, all null classifiers are payoff-equivalent to the designer, and so $\delta_\emptyset^*=(\delta, 1-\delta)$ for any $\delta\in[0, 1]$. In some notable cases, as we will prove, the only possible equilibria are null.

%Moreover, there may exist a continuum of rewards $r$ that can sustain a null equilibrium, though all are payoff equivalent to both the designer and the voter.  For simplicity, we restrict attention to null equilibria that set $r^*=0$.    %Note that a null equilibrium may be preferred by the designer to a non-null one! 
The next proposition characterizes all non-null equilibria.
\begin{proposition} 
\label{theEquilibria}
In any non-null equilibrium, $r^*={k_\mu^*\over{(\delta_1^*+\delta_0^*-1)(2\phi-1)}}$  and $\delta^*$ is as follows: 
\begin{itemize}
\item If $k_\mu^*(t,F)>0$, 
\[
\delta^*=\begin{cases}
(1, \delta_0^*(k_\mu^*, 1)) \text{ with }\delta_0^*(k_\mu^*(t,F), 1)\not=0, \text{ or }\\
 (\delta_1^*(k_\mu^*(t,F), 0), 0) \text{ with }\delta_1^*(k_\mu^*(t,F), 0)\not=1
\end{cases}
\]

\item If $k_\mu^*(t,F)<0$, 
\[\delta^*=\begin{cases}
(0, \delta_0^*(k_\mu^*(t,F), 0)) \text{ with }\delta_0^*(k_\mu^*(t,F), 0)\not=1,\text{ or }\\
 (\delta_1^*(k_\mu^*(t,F), 1), 1)\text{ with }\delta_1^*(k_\mu^*(t,F), 1)\not=0.
\end{cases}
\]
\end{itemize}
\end{proposition}
Proposition \ref{theEquilibria} establishes that there are four types of non-null equilibria, but only two are relevant in any particular setting (\textit{i.e.}, for any pair $(F,\phi)$), depending on whether the median voter wants to increase or decrease compliance relative to the sincere prevalence.\footnote{If the median voter does not want to change compliance from the sincere prevalence, then Proposition \ref{Pr:NullExistence} implies that there is an equilibrium with either $r^*=0$ or a null classifier (or both). 
%\textcolor{red}{BOOBOO We need to be careful here. There might not be an equilibrium with a NULL CLASSIFIER but there will be one with an $r=0$ and $D$ optimizing in light of that. So we will have equilibrium existence, it just might be with a perfectly responsive classifier of $(1, 1)$ if that makes sense.} \textcolor{blue}{JWP: Yes, makes sense, thanks.  Does this read better?}
}  In each case, there may exist one equilibrium with a positively responsive algorithm and/or one equilibrium with a negatively responsive algorithm.  One important aspect of this result from a substantive standpoint is that there cannot exist multiple non-null equilibria with algorithms that have the same form of responsiveness.  

Another important implication of Proposition \ref{theEquilibria} is that there may exist an equilibrium with a negatively responsive algorithm even when the median voter wants to increase compliance (\textit{i.e.}, $k_\mu^*(t,F)>0$) and, similarly, there may exist an equilibrium with a positively responsive algorithm when the median voter wants to decrease compliance (\textit{i.e.}, $k_\mu^*(t,F)<0$).  This implies that the \textit{sign} of the reward in equilibrium might apparently contradict the median voter's preference regarding compliance.  For example, it is possible for the equilibrium to involve \textit{negative} rewards ($r<0$) even if the median voter wants to \textit{increase} compliance above the level of sincere prevalence.  Such equilibria are admittedly strange --- in this case, a negative reward would be in equilibrium only when paired with a negatively response algorithm.  This is due to the duality of the responsiveness of the algorithm and the sign of the reward from the individuals' standpoints when choosing their behaviors, and is something we will discuss in a subsequent example.

Our final non-existence result focuses on the alignment between the designer's preferences, $(A_0,A_1,B_0,B_1)$, and the median voter's preferences about compliance ($k_\mu^*(t,F)$).  
\begin{proposition}
\label{null}
If $k_\mu^*(t,F)>0$  then there does not exist a non-null equilibrium when $A_1=B_0$, or when $B_1=B_0>A_1=A_0$. If $k_\mu^*(t,F)<0$  then there does not exist a non-null equilibrium when $A_0=B_1$, or when $A_1=A_0>B_1=B_0$.
\end{proposition} 
Proposition \ref{null} characterizes some scenarios in which the preferences of the median voter and the algorithm designer are directly opposed, in terms of the prevalence of qualification.  At a non-null equilibrium the median's preference can be characterized by whether $k_\mu^*(t,F)>0$ or $k_\mu^*(t,F)< 0$: whether the median prefers to induce \textit{greater} compliance than would attain in the absence of a reward-based classifier (i.e., $F(0)$), or whether to induce \textit{less} compliance.  When $A_1=A_0=A$ and when $B_1=B_0=B$ then the designer's preference is also fully characterized by whether he wants more or less compliance (when $A>B$ he wants more, and when $B>A$ he wants less). Consequently, when $k_\mu^*(t,F)>0$ but $B>A$ then the median voter and the designer have opposed preferences, with the median choosing an $r$ to bolster compliance and $D$ choosing a classifier to reduce compliance. As in the game of matching pennies, there is no pure strategy equilibrium. We're left with only the possibility of a null equilibrium which, unfortunately, may also fail to exist, as discussed in the next remark.

\begin{remark}
A (pure strategy) equilibrium may not exist in our framework for two reasons. The first is surmountable, and stems from the fact that the set of rewards is not bounded.  However, even if we bound the rewards the best response correspondence for the designer may not be convex-valued, and this can lead to equilibrium non-existence. Suppose that the designer highly values aggregate non-compliance, while the voters highly value aggregate compliance (setting $t$ high).  We can construct an example in which no non-null equilibrium is possible. As in a game of matching pennies, if $r>0$ then $D$ will choose a negatively responsive algorithm, which will induce the voters to choose $r<0$, which will induce $D$ to choose a positively responsive algorithm, which will induce the voters to choose $r>0$.  At the same time, a null equilibrium will not be possible for certain values of $F(0)$, as characterized in Proposition \ref{Pr:NullExistence}. This said, all examples derived in the article (even those that don't correspond to an existence result) are indeed equilibria! 
\end{remark}

\subsection{Social Welfare \label{Sec:SocialWelfare}}

In addition to considering democratically-chosen rewards and penalties, it is natural to think about the social welfare-maximizing system of rewards and penalties. Given our assumption that individuals have linear preferences over rewards and the imposition of budget balance, all wins and losses from classification are canceled out when considering Benthamite social welfare. Social welfare is calculated as the following:
\[
SW(r)=tF(r\cdot\rho(\delta, \phi))-\int_{-\infty}^{r \cdot \rho(\delta, \phi)}\gamma dF(\gamma).
\]
The following proposition neatly characterizes the social welfare optimizing reward level, given any non-null classifier $\delta$ and precision $\phi$.\footnote{When the classifier is null, all rewards are equivalent from a social welfare standpoint.}
\begin{proposition}
    \label{Pr:SocialWelfareOptimum}
    For any precision $\phi \in \left(\frac{1}{2},1\right]$ and non-null classifier $\delta$, the social welfare maximizing reward, $r^*_{SW}(\delta,\phi)$ is defined by:
    \begin{equation}
    \label{Eq:SocialWelfareOptimalReward}
        r^*_{SW}(\delta,\phi)=\frac{t}{\rho(\delta, \phi)}.    
    \end{equation}
\end{proposition}
Proposition \ref{Pr:SocialWelfareOptimum} implies that, while democratically-chosen rewards are a function of the overall distribution of costs, $F$,  social welfare-maximizing rewards are not. Finally, we note the following corollary of Proposition \ref{Pr:SocialWelfareOptimum} and Corollary \ref{eqRCorr}:
\begin{corollary}
For any precision $\phi \in \left(\frac{1}{2},1\right]$ and classifier $\delta$, the social welfare maximizing reward, $r^*_{SW}(\delta,\phi)$, is strictly between the optimal individual rewards conditional on non-compliance and compliance ($r_o^*$ and $r_1^*$, respectively):
\[
r_0^*< r_{SW}^*< r_1^*.
\]
\end{corollary}
Consequently, \textit{democratically chosen rewards are always inconsistent with social welfare maximization, being either lower or higher than socially optimal}.  This is in line with many other models of democratic choice (including \cite{MeltzerRichard81}, upon which many such models are based).

\subsection{An Aside on Voter Preferences and Virtual Values}

The interested reader may note that the expressions for $k_1$ and $k_0$ presented in Corollary \ref{eqRCorr} bear a resemblance to virtual valuations in Bayesian mechanism design.  This is not a coincidence, and in this section we briefly lay out the relationship between virtual values and optimal rewards, from the voter's perspective.  First, note that the expected ``profit" to choosing $\beta_i=1$ over $\beta_i=0$ is represented by $r\cdot \rho(\delta, \phi)$, and the cost of this choice is $\gamma_i$.  

 A voter who has chosen $\beta_i=1$ over $\beta_i=0$ faces an objective function given in Equation \ref{rInvest}: $$-\gamma_i+r\cdot \rho(\delta, \phi)(1-F(r\cdot \rho(\delta, \phi)))+tF(r\cdot \rho(\delta, \phi)).$$ The middle term $r\cdot \rho(\delta, \phi)(1-F(r\cdot \rho(\delta, \phi)))$ is identical to the objective of a profit-maximizing mechanism designer (e.g. a firm) who faces a buyer with value for a good that is distributed according to $F$.  The designer seeks to set a price $r\cdot \rho(\delta, \phi)$ to maximize expected revenue, which of course is $r\cdot \rho(\delta, \phi)$ times the probability the buyer's valuation for the good exceeds its price, or $1-F(r\cdot \rho(\delta, \phi))$.  In our setting, the compliant voter seeks to maximize the expected payoff a compliant type will receive conditional on budget balance. This payoff is increasing in $r\cdot \rho(\delta, \phi)$, but decreasing in $F(r\cdot \rho(\delta, \phi))$, or the set of compliant individuals expected to receive the payoff.  When $t=0$, the solution to the compliant voter's problem sets $$r\cdot \rho(\delta, \phi)-{{1-F(r\cdot \rho(\delta, \phi))}\over{f(r\cdot \rho(\delta, \phi))}}=0,$$ which is precisely the condition of choosing $r\cdot \rho(\delta, \phi)$ to set the virtual value of a voter with costs $\gamma$ distributed $F(\gamma)$ equal to zero. %Note that virtual value is basically marginal revenue.
 
Conversely, a voter who has chosen $\beta_i=0$ over $\beta_i=1$ faces an objective function given in Equation \ref{rNoInvest}, which we can reduce to $$-r\cdot \rho(\delta, \phi)(F(r\cdot \rho(\delta, \phi)))+tF(r\cdot \rho(\delta, \phi)).$$ In this case, $-r\cdot \rho(\delta, \phi)$ represents the expected profit to choosing $\beta_i=0$ over $\beta_i=1$. When $t=0$ the non-compliant voter simply seeks to maximize the expected payoff a non-compliant type will receive conditional on budget balance, or $-r\cdot \rho(\delta, \phi)(F(r\cdot \rho(\delta, \phi)))$. This term is increasing in $-r\cdot \rho(\delta, \phi)$ and increasing in $F(r\cdot \rho(\delta, \phi))$, the fraction of compliant types. In this case, with $t=0$, the non-compliant voter optimally sets $$r\cdot \rho(\delta, \phi)+{{F(r\cdot \rho(\delta, \phi))}\over{f(r\cdot \rho(\delta, \phi))}}=0.$$

\subsection{Returning to accuracy minimization}

We now return to the example of accuracy minimization that we considered in Section \ref{accNeutral}. This example sets precision $\phi={3\over 4}$ and considers two different distributions of costs. We'll first let $\gamma$ be distributed $N[0, 1]$, and then we will let $\gamma$ be distributed $N[1, 1]$. In both cases we'll set the externality of compliance at $t=0.5$.  Note that we don't pin down $r$ because it is now chosen endogenously.

When costs are distributed $N[0, 1]$ the median voter prefers $k_1$ to $k_0$, setting $k_\mu^*(t,F)=1.12$.  Consequently, a non-null classifier will yield a prevalence of $F(k_\mu^*(t,F))=87\%$, whereas a null classifier will yield a prevalence of $F(0)=50\%$.  There is a unique, non-null, equilibrium:

\small
\begin{center}
 \begin{tabular}{|c|c|c|c|c|c|c|}  \hline
 \multicolumn{7}{|c|}{\textbf{Accuracy}} \\ \hline
$\delta^*_1$& $\delta^*_0$ & Reward & Prevalence, $\pi$ & Welfare &Median Payoff& Designer Payoff\\ \hline \hline 
1 &0.67&3.14&87\%&0.65 & 0.81& 0.79 \\\hline
 \end{tabular}

\end{center}

\normalsize

\ \\We'll now change our example to shift the mean of the cost distribution to $\mu=1$, keeping all else equal.  With these higher costs the median voter prefers $k_0$ to $k_1$, setting $k_\mu^*(t,F)=-0.12$ to disincentivize compliance. A non-null classifier will yield a prevalence of $F(k_\mu^*(t,F))=13\%$, whereas a null classifier will yield a prevalence of $F(0)=16\%$.  There are now \textit{three} equilibria:
\small
\begin{center}
 \begin{tabular}{|c|c|c|c|c|c|c|}  \hline
 \multicolumn{7}{|c|}{\textbf{Accuracy}} \\ \hline
$\delta^*_1$& $\delta^*_0$ & Reward & Prevalence, $\pi_F$ & Welfare &Median Payoff& Designer Payoff\\ \hline \hline 
0 &0.96& 6.02&13\%&0.15  & 0.16&0.844  \\\hline
0.18 &1&-1.34  &13\%& 0.15& 0.16& 0.85 \\\hline
0 &1& 0&16\%& 0.16 & 0.08&0.841  \\\hline
 \end{tabular}
 
\end{center}

\normalsize

This example highlights situations in which there are a multiplicity of equilibria, and why. There are always a total of two possible non-null equilibria: one corresponding to $\delta_1\in\{0, 1\}$ and $r>0$ and one corresponding to $\delta_0\in\{0, 1\}$ and $r<0$.  In this particular example, $k_\mu^*(t,F)<0$ as the median wants to negatively induce compliance.  Therefore, if $r^*>0$ then it must be that $\delta_1=0$ and if $r^*<0$ it must be that $\delta_0=1$.  We have two local optima corresponding to these points, and we can check that they are global optima by checking to ensure that $D$ doesn't receive a higher payoff with a null classifier, or at a different local optimum that is not a possible equilibrium, because it is not consistent with the median's optimal choice of $r^*$.  In this example, both potential local optima are global optima.  Moreover, there is also a null equilibrium, because at $r^*=0$ it is optimal for $D$ to choose a null classifier. 

We'll finish by briefly discussing the two non-null equilibria: $\delta^*=(0, 0.96), r^*=6.02$, and $\delta^*=(0.18, 1), r^*=-1.34$.  In both cases, individuals are being negatively incentivized to comply, but through different mechanisms.  In the former, the reward to being classified as a ``1" is positive, but the designer is more likely to give this reward to individuals sending a signal of non-compliance, $s_i=0$.  In the latter, the designer is more likely to reward individuals sending a signal of compliance, but the reward is negative---the ``reward" is actually a penalty.  %BOOBOO Language here about why this is interesting, and something we want to think more about.

\subsection{Inefficient democratic choice}
We conclude with an example showing that democratizing the system of rewards and penalties can sometimes produce pathological outcomes. In particular, there may exist an exogenously fixed reward and penalty scheme that leaves the designer and the median voter strictly better off than they are at the democratically chosen (equilibrium) system of rewards and penalties.  Moreover, this exogenous system of rewards and penalties also improves aggregate social welfare relative to the equilibrium system of rewards and penalties.  This Pareto improvement for the designer and median voter can occur if (and only if) there exists no non-null equilibrium.  This is because the median is attaining her highest possible payoff at any non-null equilibrium.
 
Suppose that costs to compliance, $\gamma$, are distributed $N[1, 1]$, that accuracy $\phi={3\over 4}$, and that the externality of compliance is set at $t=1.25$.  In this case the median voter's costs are less than $t$, and her ideal reward induces an aggregate level of compliance equal to $F(k_1)$. As $k_1=1.93$, equilibrium compliance at a non-null equilibrium is $0.82$, and for any non-null classifier, rewards are democratically set at: $$r={{1.93}\over{(\delta_1+\delta_0-1)(.5)}}.$$For any null classifier, equilibrium compliance is $F(0)=0.16.$

\ \\Now suppose that the designer has payoffs as represented in the following confusion matrix:   \begin{center}

\ \\  \begin{tabular}{|c||c|c|} \hline
&\multicolumn{2}{c|}{Decision} \\ \hline
Compliance&$d_i=1$&$d_i=0$ \\ \hline \hline

\multirow{2}{*}{$\beta_i=1$}&$A_1=1$&$A_0=0$ \\ 
& (True Positive)&(False Negative) \\ \hline
\multirow{2}{*}{$\beta_i=0$}&$B_0=0$&$B_1={9\over 10}$\\ 
& (False Positive) & (True Negative) \\ \hline
 \end{tabular}
  \end{center}
 
\ \\ The designer is accuracy-motivated, but receives a slightly higher payoff for true positives (rewarded compliers) than true negatives (penalized non-compliers).  In this case, the classifier $(\delta_1, \delta_0)=(1, 0.9)$ induces a (democratically-chosen) reward of $r=4.24$.  When $r=4.24$ then $(\delta_1, \delta_0)=(1, 0.9)$ is also a local maximum of the designer's payoff function, yielding the designer an expected payoff of $0.74$.  However, it is not a global maximum of the designer's payoff function.  If the designer chooses a null classifier of $(\delta_1, \delta_0)=(0, 1)$ (classifying each individual as a $d_1=0$) he can attain a payoff of $0.84*0.9=0.757$, or $B_1$ times the fraction of non-compliers.  Consequently, the unique equilibrium is null, with $(\delta^*_1, \delta^*_0)=(0, 1)$ and $r^*=0$. The median receives a payoff of $t*F(0)=0.2$, and aggregate social welfare is $$t\cdot F(0)-\int_{-\infty}^0 \gamma dF(\gamma)=0.28.$$

\ \\Now suppose that $r$ is increased to $r=5$.  In this case, the designer's optimal classifier is $(\delta_1, \delta_0)=(1, 0.84)$, yielding the designer a slightly higher expected payoff than what he would attain at a null classifier ($0.76$ versus $0.757$).  This classifier and reward yield a compliance rate of $\pi_F=86\%$. The median voter now receives an expected payoff of 
\[
t\cdot \pi_F(\delta,\phi,r) - \gamma_\mu + r \cdot \rho(\delta, \phi)(1-\pi_F)=0.37,
\]
and aggregate social welfare is 
\[
t\cdot \pi_F(\delta,\phi,r)-\int_{-\infty}^{r \rho(\delta, \phi)} \gamma \; \mathrm{d}F(\gamma)=0.43.
\]
The table below summarizes the comparison between outcomes in equilibrium versus outcomes when rewards and penalties are no longer endogenously chosen.

\begin{center}
\begin{tabular}{|c|c||c|c|c|c|} \hline

\multicolumn{6}{|c|}{\textbf{Equilibrium outcomes}} \\\hline
$r$&$(\delta_1, \delta_0)$&Designer payoff &Median payoff& Social Welfare& Compliance \\ \hline
$0$&$(0, 1 )$& $0.757$& $0.2$ & $0.28$& 16\% \\ \hline
\end{tabular}
\end{center}

\begin{center}
\begin{tabular}{|c|c||c|c|c|c|} \hline

\multicolumn{6}{|c|}{\textbf{Outcomes when rewards are exogenous}} \\\hline
$r$&$(\delta_1, \delta_0)$&Designer payoff &Median payoff& Social Welfare& Compliance \\ \hline
$5$&$(1, 0.84)$& $0.76$& $0.37$ & $0.43$& 86\% \\ \hline
\end{tabular}
\end{center}
\ \\The logic behind that example is that the median voter most prefers a level of compliance equal to 82\%. For any non-null classifier she will design a system of rewards to bring compliance to this level.  However, this level of compliance is too low for it to be profitable to the (accuracy-motivated) designer to use a non-null classifier. When the designer uses a null classifier he is able to (correctly) classify 84\% of the population as non-compliant.  However, by fixing a reward that is higher than the median prefers, the designer is induced to mobilize greater compliance---higher than that demanded by the median.  This benefits the designer, because he prefers correctly classifying compliers to correctly classifying non-compliers.  It also benefits the median voter and it improves aggregate social welfare, due to the positive externalities associated with compliance.

\section{Conclusion}

Classification algorithms often do more sort than simply categorize people --- they also often change peoples' behaviors. Indeed, such behavioral changes are sometimes an explicit goal of the algorithm, just as crime prediction algorithms may be designed to deter crime.  However, regardless of whether behavioral changes are the goal of algorithm designer, individuals' preferences over how they are classified by an algorithm may induce these individuals to change their behaviors.  When an algorithm affects the behaviors of the individuals to whom the algorithm is applied, the result is \textit{algorithmic endogeneity}. 

Such endogeneity accentuates the importance of the goals of the algorithm designer.  To see this, consider two similar cities, $X$ and $Y$, designing a ``ticketing algorithm'' that chooses which drivers to penalize for unsafe driving.  Suppose that city $X$'s algorithm has been designed to maximize ticket revenue, while city $Y$'s algorithm was designed to maximize public safety.  Even though each city is using an algorithm aimed at managing unsafe driving behavior, the two algorithms might in general make very different classification decisions and, as a result of algorithmic endogeneity,  driving behaviors, revenues, and/or public safety might vary widely between the two otherwise similar cities.

Accordingly, our theory provides another view on structural inequality, emanating from the incentives of those who design the algorithms applied to individuals.  This is one reason that the \textit{stakes} of algorithmic classification --- housing eligibility, pretrial release, educational opportunities, to name a few --- have been increasingly subject to scrutiny and reform.  These ongoing debates might at first appear to be about the nature of the algorithms and/or the data on which they are ``trained,'' but the analysis above indicates that, from a social science standpoint, the policy implications of these algorithms must ultimately focus on the \textit{decisions} that the algorithm in question is used to make.  

\paragraph{Our Argument.} In the analysis above, we first characterized the optimal classification algorithm for any given algorithm designer's preferences over both how people behave and how they are classified, holding the rewards and penalties individuals experience from classification fixed.  We then show that even seemingly ``neutral'' goals such as \textit{accuracy maximization} can produce much less benign outcomes in the presence of algorithmic endogeneity.  Indeed, as the stakes of algorithmic classification become sufficiently strong, any algorithm designer can induce essentially everyone to engage in any given behavior and also be classified in any fashion the designer wants.

The next step of our analysis above --- motivated by recent reforms intending to democratize the stakes of classification --- characterizes what classification algorithms will look like in equilibrium when the algorithm's stakes are subject to democratic control, subject to a budget balance condition.  The analysis shows that, for any non-null classifier, equilibrium classification algorithms induce a fixed level of behavioral compliance.  This level of compliance is optimal for the median voter, but also socially inefficient.  In addition, the median vter's ability to set the stakes so as to maintain a given level of compliance in the population dramatically limits the algorithm designer's ability to shape behavior in the population as a whole.  In the end, the algorithm designer is essentially faced with a choice between either designing a non-null classification algorithm that induces the median voter's ideal level of compliance, or a ``null" algorithm that classifies individuals randomly. 

And, in some cases, the equilibrium algorithm is random: \textit{when the preferences of voters and the designer are sufficiently opposed with respect to optimal aggregate behavior, the equilibrium algorithm must be random in the sense of being a null classifier}.  From a substantive standpoint, such null classifiers are effectively ``defunded algorithms'' because they have no impact on individual incentives.  In line with recent discussions about reducing the stakes of classification algorithms, these nnull algorithms emerge precisely in settings in which the median voter and algorithm designer have a fundamental disagreement over how social behavior should be structured.

\paragraph{Future Directions.}  There are many ways to expand the framework presented in this article.  In addition to considering richer settings (\textit{i.e.}, larger sets of behaviors and/or decisions, different informational structures, different individual preferences), an important question raised by the analysis is how to judge the \textit{fairness} of equilibrium algorithms.  The analysis above illustrates that democratically chosen stakes to classification are socially inefficient, suggesting immediately that the equilibrium algorithm is always suspect from a welfare-based fairness perspective.  

However, this merely scratches the surface of bigger questions about algorithmic fairness.  For example, if the population is divided into two or more \textbf{groups}, it is known that any algorithm is ``generically unfair'' from a statistical parity sense (\textit{e.g.}, \cite{KleinbergMullainathanRaghavan16}, \cite{Chouldechova17}).  The analysis above demonstrates that any statistical imbalance might be leveraged by the median voter when choosing the algorithm's stakes.  And, even setting democratic choice to the side, the fact that the algorithm will shape individual incentives and produce algorithmic endogeneity both raises questions about the proper definition of fairness in such situations and, indeed, opens some angles with respect to how to evaluate existing statistical notions of fairness.\footnote{Some of these issues are addressed in \cite{PattyPenn23AlgorithmicEndogeneityFairness}.} 

\newpage

\newpage

\section{Appendix}
\subsection{Equilibrium characterization \label{eqAppendix}}
\ \\These are the (potentially interior) solutions for $\delta_0^*$ and $\delta_1^*$ for our general equilibrium. They are each defined in terms of the functions $\Delta_1(k, \delta_0)$ and $\Delta_0(k, \delta_1)$, where $\Delta_i(k, \delta_j)$ solves $${{\partial }\over{\partial \delta_i}}EU_D(\delta_i, \delta_j)=0.$$ Consequently, holding $\delta_j$ and $k$ fixed, $\Delta_i(k, \delta_j)$ is the (unique) critical point of the designer's payoff function in $\delta_i$.

$$\delta_0^*(k, \delta_1)=\begin{cases}
0 \,\,\,\,\text{ if }\Delta_0(k, \delta_1)<0\\
\Delta_0(k, \delta_1) \text{ if }0\leq \Delta_0(k, \delta_1)\leq 1\\
1\,\,\,\,\text{ if }\Delta_0(k, \delta_1)>1,
\end{cases}$$ and

$$\delta_1^*(k, \delta_0)=\begin{cases}
0 \,\,\,\,\text{ if }\Delta_1(k, \delta_0)<0\\
\Delta_1(k, \delta_0) \text{ if }0\leq \Delta_1(k, \delta_0)\leq 1\\
1\,\,\,\,\text{ if }\Delta_1(k, \delta_0)>1,
\end{cases}$$ with

$$\begin{multlined}\Delta_0(k, \delta_1)={{k  f(k) ((1- {\delta_1}) \phi (- {A_0}+ {A_1}+ {B_0}- {B_1})- {A_1}+ {\delta_1} ( {B_0}- {B_1})+ {B_1})}\over{(1-\phi) ( {A_0}- {A_1}) \left(k  f(k)+F(k)\right)+\phi ( {B_1}- {B_0}) \left(1-k  f(k)-F(k)\right)}}\\
+{{(1- {\delta_1}) ((1-\phi) ( {A_0}- {A_1}) F(k)+\phi ( {B_1}- {B_0}) (1-F(k)))}\over{(1-\phi) ( {A_0}- {A_1}) \left(k  f(k)+F(k)\right)+\phi ( {B_1}- {B_0}) \left(1-k  f(k)-F(k)\right)}}
\end{multlined}
$$

$$\begin{multlined}\Delta_1(k, \delta_0)={{k f(k) ((1-{\delta_0}) \phi (-{A_0}+{A_1}+{B_0}-{B_1})+{\delta_0} ({A_1}-{A_0})-{A_1}+{B_1})}\over{\phi ({A_1}-{A_0}) \left(k f(k)+F(k)\right)+(1-\phi) ({B_0}-{B_1}) \left(1-k f(k)-F(k)\right)}}\\
+{{(1-{\delta_0}) (\phi ({A_1}-{A_0}) F(k)+(1-\phi) ({B_0}-{B_1}) (1-F(k)))}\over{\phi ({A_1}-{A_0}) \left(k f(k)+F(k)\right)+(1-\phi) ({B_0}-{B_1}) \left(1-k f(k)-F(k)\right)}}.
\end{multlined}$$
\newpage

\appendix

\section{Proofs \label{Sec:Proofs}}

\noindent \textbf{Proposition \ref{allCells}}
\textit{As $r\rightarrow\infty$ the designer can attain an expected payoff of $\max(\{A_1, A_0, B_1, B_0\})$. This is his highest possible payoff.}

\begin{proof} First note that, as $r\rightarrow\infty$, prevalence $\pi_F(\delta,\phi,r) \rightarrow 1$ if $\delta_1+\delta_0>1$ and prevalence $\pi_F(\delta,\phi,r)\rightarrow 0$ if $\delta_1+\delta_0<1$, by Equations \ref{Eq:GammaEq} and \ref{Eq:Equilibrium}. (When $\delta_1+\delta_0=1$ our classifier is null, and $\pi_F=F(0)$.)

The fraction of individuals classified into each cell of the designer's payoff matrix is a function of $\pi_F, \phi,$ and $\delta=(\delta_1, \delta_0)$, with:
\[
\begin{array}{rcl}\%A_1&=&\pi( \phi\delta_1+(1-\phi)(1-\delta_0))\\
\%A_0&=&\pi(\phi(1-\delta_1)+(1-\phi)\delta_0)\\
\%B_1&=&(1-\pi)(\phi\delta_0+(1-\phi)(1-\delta_1))\\
\%B_0&=&(1-\pi)(\phi(1-\delta_0)+(1-\phi)\delta_1).\\
\end{array}
\]

\noindent For $\epsilon$ small and positive, consider the classifiers $\delta_{A_1}=(1, \epsilon)$, $\delta_{A_0}=(\epsilon, 1)$, $\delta_{B_1}=(0, 1-\epsilon)$, and $\delta_{B_0}=(1-\epsilon, 0)$.  Classifiers $\delta_{A_1}$ and $\delta_{A_0}$ induce a prevalence near 1 when $r$ is high, and classifiers $\delta_{B_1}$ and $\delta_{B_0}$ induce a prevalence near 0.  Evaluating the above four equations at these respective classifiers, we can see that as $r\rightarrow\infty$,  classifier $\delta_{C}$ induces a $\%C=1$, for $C\in\{A_1, A_0, B_1, B_0\}$. 
\end{proof}

\noindent \textbf{Proposition \ref{Pr:OptimalComplianceAlgorithm}}
    \textit{
    If the designer's preferences $\eta$ satisfy $A_1=A_0=\bar{A}\geq 0$ and $B_1=B_0=\bar{B}\geq 0$, then the optimal algorithm for any $r>0$ is }
    \[
    \delta^*(r,F,\phi,\eta) = \begin{cases}
        (1,1) & \text{ if } \;\; r\cdot (\bar{A}-\bar{B}) > 0, \\
        (0,0) & \text{ if } \;\; r\cdot (\bar{A}-\bar{B}) < 0.
    \end{cases}
    \]
\begin{proof} 
Fixing $\phi \in (1/2,1]$ and $r > 0$, noting that $A_1=A_0>0$ by hypothesis, and normalizing $B_1=B_0=0$, Equation \eqref{Eq:ObjectiveD} can be rewritten as 
\[
EU_D(\delta \mid r,F,\phi,\eta) = \pi_F(\delta,\phi,r) \cdot A_0.
\]
Because $A_0>0$ and $\pi_F(\delta)$ is maximized by $\delta=(1,1)$ if $r>0$, it follows that $r>0$ implies that $EU_D(\delta)$ is maximized by $\delta=(1,1)$, as claimed. A similar argument proves the case of $r<0$.
\end{proof}

\noindent \textbf{Proposition \ref{Pr:Corner}} \textit{When $r\not=0$, any optimal classification strategy for $D$ requires either $\delta_1^*\in \{0, 1\}$, or $\delta_0^*\in\{0, 1\}$, or both. %When $\phi=0.5$, this holds \emph{unless}
%\[
%F(0)A_1+(1-F(0))B_0=F(0)A_0+(1-F(0))B_1,
%\]
%in which case any classification strategy is optimal for $D$. 
}
\begin{proof} The determinant of the Hessian of $D$'s objective function (a function of $\delta_1$ and $ \delta_0$) is: 
\[
\bigg| \; H(EU_D(\delta_1, \delta_0 \mid r,F,\phi,\eta)) \; \bigg|=-(A_0 - A_1 + B_0 - B_1)^2 \cdot r^2 \cdot (1 - 2 \phi)^2 \cdot F^\prime(\rho(\delta,\phi))^2.
\]
 This is strictly negative if $\phi>0$ and $ A_1+B_1\not=A_0+B_0$ whenever $F^\prime (r \left(\delta_1+\delta_0-1\right)(2\phi-1))>0$, which is implied by Assumption \ref{cdfAss}. Consequently,  $\delta_1^*, \delta_0^*$ are not both interior when $A_1+B_1\not=A_0+B_0$.  
If $A_1+B_1=A_0+B_0$ then:

$${{\partial EU_D}\over{\partial \delta_1}}=(B_0-B_1)(1-\phi+(2\phi-1)F(r\cdot \rho(\delta, \phi)))+r(2\phi-1) f(r\cdot\rho(\delta, \phi))(A_0-B_1+(B_0-B_1)\rho(\delta, \phi)),$$

$${{\partial EU_D}\over{\partial \delta_0}}=(B_0-B_1)(-\phi+(2\phi-1)F(r\cdot \rho(\delta, \phi)))+r(2\phi-1) f(r\cdot\rho(\delta, \phi))(A_0-B_1+(B_0-B_1)\rho(\delta, \phi)).$$ Any critical point in $(\delta_1, \delta_0)$ would set the above expressions equal to zero, and consequently require that $1-\phi=-\phi$, which is impossible.  Consequently, when $A_1+B_1=A_0+B_0$, $D$'s payoff has no interior critical point. 
\end{proof}

\paragraph{A Note on Log-Concavity.} Much of our analysis utilizes the log-concavity of the PDF $f$.  First, log-concavity of  $f$ implies that its CDF, $F$ is also log-concave. This implies that at any $\gamma\in \mathbf{R}$, it is the case that $f(\gamma)^2\geq F(\gamma)\cdot f^\prime(\gamma)$.  Moreover, if density $f$ is log-concave then its survival function, $1-F$ is also log-concave.\footnote{Bagnoli \& Bergstrom, ``Log-concave probability and its applications," \textit{Economic Theory} 26(2), 2005.} This leads to the following observation:

\begin{observation} When PDF $f$ is log-concave then for any $\gamma\in \mathbf{R}$ it is the case that  $$f(\gamma)^2\geq -f^\prime(\gamma)(1-F(\gamma)).$$ \label{LCobs}
\end{observation}

\noindent \textbf{Proposition \ref{bigConcavity}}
\textit{When $[A_1\geq A_0 \, \&\, B_1\geq B_0]$ with one inequality strict, or when  $[A_1\leq A_0 \, \& \, B_1\leq B_0]$ with one inequality strict then $D$'s payoff is strictly quasiconcave in $\delta_j$ and strictly quasiconvex in $\delta_k$, for $j\not=k$. }
\begin{proof} 
Our proof proceeds by considering the concavity and convexity properties of any critical points of the designer's objective function.  Taking the partial derivatives of the designer's objective function (Equation \ref{Eq:ObjectiveD}) with respect to $\delta_0$ and $\delta_1$, any critical point that is interior for either $\delta_1$ or $\delta_0$ (which we will term $\delta_1^c$ and $\delta_0^c$) must, respectively, solve:

\scriptsize
$$\delta_1^c = {{(B_1 - B_0)(1-\mathbf{F_1}-\phi) + \mathbf{F_1} \phi (A_0 -A_1-B_0+B_1) - 
     \mathbf{f_1} r(2 \phi-1) (A_1-B_1+\delta_0(A_0-A_1)+\phi(1-\delta_0)(A_0 - A_1 - B_0 + B_1))} \over{\mathbf{f_1} r(2 \phi-1) (B_1 -B_0 - \phi (A_0 - A_1 + B_1-B_0) ) }},$$

$$\delta_0^c = {{(B_0 -B_1)\phi(1-\mathbf{F_0})  + \mathbf{F_0}(1-\phi) (A_1 - A_0) + 
     \mathbf{f_0} ( 2 \phi-1) r( B_1-A_1 +\delta_1(B_0-B_1)+\phi(1-\delta_1) (A_1-A_0 + B_0 - B_1))}
        \over{\mathbf{f_0} r(2 \phi-1)  ( A_0 -A_1 +\phi (A_1 -A_0+ B_0 - B_1) )}}.$$

\normalsize  \noindent The terms $\mathbf{F_1, F_0, f_1, f_0} $ are functions of $\delta_1, \delta_0$, with $\mathbf{F_1}=F(r\cdot \rho((\delta_1^c, \delta_0), \phi))$, $\mathbf{F_0}=F(r\cdot \rho((\delta_1, \delta_0^c), \phi))$, $\mathbf{f_1}=f(r\cdot \rho((\delta_1, \delta_0^c), \phi))$, and $\mathbf{f_0}=f(r\cdot \rho((\delta_1, \delta_0^c), \phi))$.

We can now classify the second order behavior of the designer's objective function at the critical points $\delta_1^c$ and $\delta_0^c$ (holding $\delta_0$ and $\delta_1$ fixed, respectively).  The second derivatives with respect to $\delta_1$ and $\delta_0$, evaluated at $\delta_1^c$ and $\delta_0^c$, are:

%This is the second derivative with respect to $\delta_1$: $${{\partial^2 EU_D(\delta_1, \delta_0 \mid r,F,\phi,\eta)}\over{\partial \delta_1^2}}= (2 h-1) r F'((2 h-1) r (\text{d0}+\text{d1}-1)) (h (\text{d0}+\text{d1}-1) (-A_0+A_1+B_0-B_1)+A_0 \text{d0}+A_1 (-\text{d0})+A_1-B_0 \text{d1}+B_1 \text{d1}-B_1)+(-h (A_0-A_1+B_1)+B_0 (h-1)+B_1) F((2 h-1) r (\text{d0}+\text{d1}-1))+(h-1) (-(B_0-B_1))$$

%This is the second derivative with respect to $\delta_0$: $${{\partial^2 EU_D(\delta_1, \delta_0)}\over{\partial \delta_0^2}}=(2 h-1) r \left((2 h-1) r F''((2 h-1) r (\text{d0}+\text{d1}-1)) (h (\text{d0}+\text{d1}-1) (-A_0+A_1+B_0-B_1)+A_0 \text{d0}+A_1 (-\text{d0})+A_1-B_0 \text{d1}+B_1 \text{d1}-B_1)+2 (h (-A_0+A_1+B_0-B_1)+A_0-A_1) F'((2 h-1) r (\text{d0}+\text{d1}-1))\right) $$

\footnotesize

\begin{equation}{{\partial^2 EU_D(\delta_1, \delta_0)}\over{\partial \delta_1^2}}\Big|_{\delta_1=\delta_1^c}={{r(2 \phi-1) \left(\phi (A_1-A_0) (2 \mathbf{f_1}^2-\mathbf{F_1} \mathbf{f_1}^\prime)+(B_1-B_0) (1-\phi) (2 \mathbf{f_1}^2+\mathbf{f_1}^\prime-\mathbf{F_1} \mathbf{f_1}^\prime )\right)}\over{\mathbf{f_1}}},\label{d1critical}\end{equation}

\begin{equation}{{\partial^2 EU_D(\delta_1, \delta_0)}\over{\partial \delta_0^2}}\Big|_{\delta_0=\delta_0^c}=-\frac{r (2 \phi-1)  \left((1-\phi) (A_1-A_0) \left(2 \mathbf{f_0}^2-\mathbf{F_0} \mathbf{f_0}^\prime\right)+\phi (B_1-B_0) \left(2 \mathbf{f_0}^2+\mathbf{f_0}^\prime-\mathbf{F_0} \mathbf{f_0}^\prime\right)\right)}{\mathbf{f_0}}.\label{d0critical}\end{equation}

\normalsize

\noindent By the full support and log-concavity of $f$, the terms $(2 \mathbf{f_j}^2-\mathbf{F_j} \mathbf{f_j}^\prime)$ and $(2 \mathbf{f_j}^2+\mathbf{f_j}^\prime-\mathbf{F_j} \mathbf{f_j}^\prime )$ are both strictly positive (note that if $\mathbf{f_j}^\prime<0$ then this conclusion holds via Observation \ref{LCobs}). Consequently, when the values of $A_1, A_0, B_1, B_0$ allow us to unambiguously sign Equations \ref{d1critical} and \ref{d0critical} we can draw conclusions about the strict quasiconcavity of $D$'s objective function in $\delta_1$ and $\delta_0$. If, for example, Equation \ref{d1critical} is always strictly positive, then holding $\delta_0$ constant, any critical point in $\delta_1$ must be a local minimum.  Consequently, $D$'s objective function must be strictly quasiconvex in $\delta_1$ for any $\delta_0$, and therefore maximized at a corner solution $\delta_1^*\in\{0, 1\}$.  This leads to a few conclusions:

\begin{itemize}
\item If $r>0$ and $A_1\geq A_0$ and $B_1\geq B_0$ with one inequality strict, or if $r<0$ and $A_1\leq A_0$ and $B_1\leq B_0$ with one inequality strict, then $\delta_1^*\in\{0, 1\}$ and $\delta_0^*$ may be interior.
\item If $r<0$ and $A_1\geq A_0$ and $B_1\geq B_0$ with one inequality strict, or if $r>0$ and $A_1\leq A_0$ and $B_1\leq B_0$ with one inequality strict,  then $\delta_0^*\in\{0, 1\}$ and $\delta_1^*$ may be interior.
\end{itemize}
\end{proof}

\noindent \textbf{Proposition \ref{intR}} \textit{If $\delta$ is not null, then conditional on behavior $\beta_i$, voter payoffs are strictly quasi-concave in rewards, $r$, and maximized at an interior $r$. If $\delta$ is null, each voter $i$ is indifferent between all reward levels.}

\begin{proof} 
First, note that if $\delta_1+\delta_0=1$ then the classifier is null, and classifies individuals independently of their signal.  In this case, voter expected payoffs are flat in $r$, as each individual receives the reward with the same probability and pays a tax equal to this expected reward.

\ \\ Now suppose that $\delta_1+\delta_0\not=1$.  Any  critical points of Equations \ref{rInvest} and \ref{rNoInvest} (we term them $r_1^c$ and $r_0^c$, respectively) must  satisfy the following first-order conditions:

\begin{equation}r_1^c={{1-F(r_1^c\cdot \rho(\delta, \phi))+t  f(r_1^c\cdot \rho(\delta, \phi))}\over{(\delta_1+\delta_0-1)(2\phi-1)f(r_1^c\cdot \rho(\delta, \phi))}},\label{r1}\end{equation} and

\begin{equation}
\label{r0}r_0^c={{-F(r_0^c\cdot \rho(\delta, \phi))+t  f(r_0^c\cdot \rho(\delta, \phi))}\over{(\delta_1+\delta_0-1)(2\phi-1)f(r_0^c\cdot \rho(\delta, \phi))}}.\end{equation}

\ \\ Evaluating the second-order conditions of Equations \ref{rInvest} and \ref{rNoInvest} at these points, we get:

$${{\partial^2 U_V(r)}\over{\partial r^2}}\Big|_{r=r_1^c}=-{1\over{f(r_1^c\cdot \rho(\delta, \phi))}}(\delta_0+\delta_1-1)^2 (2\phi-1)^2 (2f(r_1^c\cdot \rho(\delta, \phi))^2+f^\prime(r_1^c\cdot\rho(\delta, \phi))(1-F(r_1^c\cdot \rho(\delta, \phi)))),$$ and

$${{\partial^2 U_V(r)}\over{\partial r^2}}\Big|_{r=r_0^c}=-{1\over{f(r_0^c\cdot \rho(\delta, \phi))}}(\delta_0+\delta_1-1)^2 (2\phi-1)^2 (2f(r_0^c\cdot \rho(\delta, \phi))^2-f^\prime(r_0^c\cdot\rho(\delta, \phi))F(r_0^c\cdot \rho(\delta, \phi))).$$ Again, by the strict log-concavity of the PDF of the cost distribution $f$, both these terms are strictly negative when $\delta_1+\delta_0\not=1$.  Consequently, any critical point must be a local maximum.  There is therefore at most one critical point, and it is a global maximum. 

\ \\We now show that there exists a critical point for each of the above payoff functions.  We will use the intermediate value theorem to show that both Equations \ref{r1} and \ref{r0} have roots.

\ \\We'll consider Equation \ref{r1} first, and begin by assuming that $\delta_0+\delta_1>1$. When this latter condition holds then $${{1-F(r_1^c\cdot \rho(\delta, \phi))+t  f(r_1^c\cdot \rho(\delta, \phi))}\over{(\delta_1+\delta_0-1)(2\phi-1)f(r_1^c\cdot \rho(\delta, \phi))}}>0.$$  Consequently, when $r_1^c\leq 0$ then $$ r_1^c-{{1-F(r_1^c\cdot \rho(\delta, \phi))+t  f(r_1^c\cdot \rho(\delta, \phi))}\over{(\delta_1+\delta_0-1)(2\phi-1)f(r_1^c\cdot \rho(\delta, \phi))}}<0.$$ As we drive $r_1^c$ to infinity we get $$\lim_{r_1\rightarrow \infty} r_1^c-{{1-F(r_1^c\cdot \rho(\delta, \phi))+t  f(r_1^c\cdot \rho(\delta, \phi))}\over{(\delta_1+\delta_0-1)(2\phi-1)f(r_1^c\cdot \rho(\delta, \phi))}}$$
\begin{equation}=\lim_{r_1\rightarrow \infty} r_1^c-{{t }\over{(\delta_1+\delta_0-1)(2\phi-1)}}-{1\over{(\delta_1+\delta_0-1)(2\phi-1)}} \lim_{r_1^c\rightarrow \infty}{{(1-F(r_1^c\cdot \rho(\delta, \phi)))}\over{f(r_1^c\cdot \rho(\delta, \phi))}}.\label{posThing}\end{equation}  To prove that Equation \ref{posThing} is positive, it suffices to show that the term ${{(1-F(r_1^c\cdot \rho(\delta, \phi)))}\over{f(r_1^c\cdot \rho(\delta, \phi))}}$ does not approach $\infty$ as $r_1^c\rightarrow \infty$.  As noted in Observation \ref{LCobs}, log-concavity of $f$ implies log-concavity of its survival function $1-F$.  Therefore $\log(1-F)$ is a concave function.  $${d\over{d\gamma}}\log(1-F(\gamma))=-{{f(\gamma)}\over{1-F(\gamma)}}$$ must consequently be decreasing, which implies that ${1-F(\gamma)}\over{f(\gamma)}$ is also decreasing.  As ${1-F(\gamma)}\over{f(\gamma)}$ is decreasing and bounded below by $0$, it must converge to a limit as $r_1^c\rightarrow \infty$.  Consequently, by the intermediate value theorem, when $\delta_1+\delta_0>1$ Equation \ref{r1} has a root.

\ \\ Now consider Equation \ref{r1}  assuming that $\delta_0+\delta_1<1$.  In this case, when $r_1^c>0$ then $$ r_1^c-{{1-F(r_1^c\cdot \rho(\delta, \phi))+t  f(r_1^c\cdot \rho(\delta, \phi))}\over{(\delta_1+\delta_0-1)(2\phi-1)f(r_1^c\cdot \rho(\delta, \phi))}})>0.$$  By a similar argument as above, $$\lim_{r_1^c\rightarrow-\infty} r_1^c-{{1-F(r_1^c\cdot \rho(\delta, \phi))+t  f(r_1^c\cdot \rho(\delta, \phi))}\over{(\delta_1+\delta_0-1)(2\phi-1)f(r_1^c\cdot \rho(\delta, \phi))}})<0$$ by the the fact that $-{{1-F(\gamma)}\over{f(\gamma)}}$ is increasing and bounded above by $0$.  Again, by the intermediate value theorem, when $\delta_1+\delta_0<1$ Equation \ref{r1} has a root.  A similar argument proves that Equation \ref{r0} has a root when $\delta_1+\delta_0\not=1$.  Consequently, conditional on behavior $\beta_i$, the voter's objective function has a unique and interior maximizer.
\end{proof}

\noindent \textbf{Proposition \ref{symmetricDist}} \textit{
For any $t$, $F$, and $\phi$, and any voter $i \in N$,
\[
\bigg[ \; k_i^*(t,F)=k_1^*(t,F) \; \text{ \emph{and} } \; t'>t \; \bigg] \Rightarrow \;\; k_i^*(t',F)=k_1^*(t',F).
\]}

\begin{proof} 
By Equation \ref{kStar}, $k_i^*=k_1$ if and only if $\gamma_i\leq k_0\cdot F(k_0)+k_1\cdot(1-F(k_1))+t\cdot(F(k_1)-F(k_0)). $  We will show that the term \begin{equation}k_0\cdot F(k_0)+k_1\cdot(1-F(k_1))+t\cdot(F(k_1)-F(k_0))\label{muGuy}\end{equation} is increasing in $t$.

\ \\Substituting the expressions for $k_1$ and $k_0$ into  (\ref{muGuy}), we can restate our problem as needing to show that  \begin{equation}\label{muGuy2}t+{{(1-F(k_1))^2}\over{f(k_1)}}-{{F(k_0)^2}\over{f(k_0)}}\end{equation} is increasing in $t$ (and moreover, it is strictly increasing in $t$).  Using the fact that $t+{{(1-F(k_1))}\over{f(k_1)}}-k_1=0$ and $t-{{(F(k_0))}\over{f(k_0)}}-k_0=0$, we can implicitly differentiate $k_1$ and $k_0$ with respect to $t$ to yield:

$$\begin{array}{rcl} k^\prime_1(t)&=&{{f(k_1)^2}\over{2f(k_1)^2}+(1-F(k_1))f^\prime(k_1)}, \text{ and }\\
k^\prime_0(t)&=&{{f(k_0)^2}\over{2f(k_0)^2}-F(k_0)f^\prime(k_0)}.\end{array}$$ 

\ \\Finally, differentiating (\ref{muGuy2}) with respect to $t$ and substituting the above solutions for ${{d }\over{d t}}k_1$ and ${{d }\over{d t}}k_0$ into the expression, we get that $${d\over{dt}}\big( t+{{(1-F(k_1))^2}\over{f(k_1)}}-{{F(k_0)^2}\over{f(k_0)}}\big) =F(k_1)-F(k_0).$$ As $k_1>k_0$, this expression is  positive, showing that  (\ref{muGuy2}) is strictly increasing in $t$. 
\end{proof}

%\noindent \textbf{Proposition 
%\ref{Pr:FOSDOrderingOfKStar}}
%\textit{For any $t$, $F$, and $\phi$, and any voter $i \in N$,}
%\[
%\bigg[ \; k_i^*(t,F)=k_1^*(t,F) \; \text{ \emph{and} } \; F \succ F' \; \bigg] \Rightarrow \;\; k_i^*(t,F')=k_1^*(t,F').
%\]
%\begin{proof}
  %  To be added.
%\end{proof}

%BOOBOOPROOFSEND

\section{Derivations \label{Sec:TechnicalStuff}}

\subsection{The Voter's Expected Payoffs}

The conditional expected payoff function for individual $i$ in \eqref{twoPeak} is derived from the following two conditional expected payoffs conditional on $i$'s subsequent behavioral choice, $\beta_i$:
\small

\ \\$EU_V(\beta_i=1 \mid r, \gamma_i, \delta, t, F, \phi)$
\begin{eqnarray}
& = & t F(r\cdot\rho(\delta, \phi))+r (\phi \delta_1+(1-\phi)(1-\delta_0))-\gamma_i \nonumber \\
& & -r \big(F(r\cdot\rho(\delta, \phi))(\phi \delta_1+(1-\phi)(1-\delta_0))+(1-F(r\cdot\rho(\delta, \phi)))(\phi(1-\delta_0)+(1-\phi)\delta_1)\big)\nonumber \\
& & \nonumber \\
& = & -\gamma_i+r\cdot\rho(\delta, \phi)(1-F(r\cdot\rho(\delta, \phi)))+t\cdot F(r\cdot\rho(\delta, \phi)), \label{rInvest}
\end{eqnarray}

$EU_V(\beta_i=0 \mid r, \gamma_i, \delta, t, F, \phi)$
\begin{eqnarray}
& = & t F(r\cdot\rho(\delta, \phi))+r (\phi(1-\delta_0) +(1-\phi)\delta_1) \nonumber \\
& & -r \big(F(r\cdot\rho(\delta, \phi))(\phi \delta_1+(1-\phi)(1-\delta_0))+(1-F(r\cdot\rho(\delta, \phi)))(\phi(1-\delta_0)+(1-\phi)\delta_1)\big) \nonumber \\
& & \nonumber \\
& = & -r\cdot\rho(\delta, \phi) F(r\cdot\rho(\delta, \phi)) +t\cdot F(r\cdot\rho(\delta, \phi)).\label{rNoInvest}
\end{eqnarray} 

\normalsize
\noindent \textbf{Proposition \ref{Pr:MedianCutpoint}}
\textit{If $f$ is log-concave and symmetric about its mean, then the median voter ($i=\mu$) receives a higher payoff at $r_1^*$ than $r_0^*$ if and only if $\gamma_\mu\leq t$.}
\begin{proof}
    When $f$ is symmetric about mean $\mu$, then if $t=\mu$, $k_1$ and $k_0$ are symmetric about $\mu$.  To see this, let $k_1$ and $k_0$ be defined as in Corollary \ref{eqRCorr}.  If $k_1$ solves $$k_1=\mu+{{1-F(k_1)}\over{f(k_1)}}$$ then $2\mu-k_1$ must solve $$2\mu-k_1=\mu-{{F(2\mu-k_1)}\over{f(2\mu-k_1)}}.$$  This is because, by the symmetry of $f$ about $\mu$, $f(2\mu-k_1)=f(k_1)$ and $F(2\mu-k_1)=1-F(k_1)$.  Consequently, $k_0=2\mu-k_1$.  By Equation \ref{kStar}, a voter with $\gamma=\mu$ is indifferent between $k_1$ and $k_0$ if $t=\mu$, and by Proposition \ref{symmetricDist} this is the unique point at which the median voter is indifferent. $\Box$
\end{proof}

\noindent \textbf{Proposition \ref{Pr:MedianVoterTheorem}} \textit{
For any classifier, $\delta$, marginal value of compliance, $t$, distribution $F$, and precision $\phi$, the reward 
\[
r^*(\delta\mid t, F, \phi) = \frac{k_\mu^*(t,F)}{\rho(\delta, \phi)},
\]
is a Condorcet winner: it is preferred by a majority of individuals to \textit{any} other reward, $r \in \mathbf{R}$.}
 \begin{proof} 
 By Equation \ref{kStar}, individuals with $\gamma_i>k_0\cdot F(k_0)+k_1\cdot(1-F(k_1))+t\cdot(F(k_1)-F(k_0))$ have ideal point $$r_i^*={{k_0}\over{\rho(\delta, \phi)}}$$ and individuals with $\gamma_i\leq k_0\cdot F(k_0)+k_1\cdot(1-F(k_1))+t\cdot(F(k_1)-F(k_0))$ have ideal point $$r_i^*={{k_1}\over{\rho(\delta, \phi)}}.$$  Consequently, a majority of individuals will share the median voter's ideal point in $r$. 
 \end{proof}

\noindent \textbf{Proposition \ref{Pr:VoterIndifferenceNonNull}}
\textit{When rewards are chosen democratically, every voter is indifferent between all non-null classifiers. Regardless of how rewards are chosen, every voter is indifferent between all null classifiers.  }

\begin{proof} By Corollary \ref{eqRCorr} we know that, given any non-null classifier and at rewards $r^*$, individuals will choose $\beta_i=1$ if and only if  $\gamma_i\leq k_\mu^*(t,F)$, and that $k_\mu^*(t,F)$ is invariant to any non-null classifier.  Substituting $r^*$ into Equations \ref{rInvest} and \ref{rNoInvest}, individuals choosing $\beta_i=1$ receive a payoff of $k_\mu^*(1-F(k_\mu^*))+t F(k_\mu^*)-\gamma_i$ and individuals choosing $\beta_i^*=0$ receive a payoff of $tF(k_\mu^*)-k_\mu^*(t,F) \cdot F(k_\mu^*)$.  Both these payoffs are invariant to classifier, and as $k_\mu^*(t,F)$ is constant across non-null classifiers, every voter is indifferent over every non-null classifier when rewards are democratically chosen.

For a null classifier, $\delta_1=1-\delta_0$.  Substituting this into Equations \ref{rInvest} and \ref{rNoInvest}, individuals with $\gamma_i\leq 0$ receive a payoff of $t F(0)-\gamma_i$, and individuals with $\gamma_i>0$ receive a payoff of $t F(0)$.  Again, payoffs are invariant to classifier, for any null classifier. 
\end{proof}

\noindent \textbf{Proposition \ref{Pr:NullExistence}}  \textit{When $k_\mu^*(t,F)\not=0$, there never exists a null equilibrium when: $$F(0)\in \left( {{(B_1-B_0)(1-\phi)}\over{B_1-B_0-\phi(A_0-A_1+B_1-B_0)}},{{(B_1-B_0)\phi}\over{A_1-A_0-\phi(A_1-A_0-B_1+B_0)}}\right).$$ Otherwise, there always exists a null equilibrium.}
\begin{proof} 
Note that when $A_1=A_0$ and $B_1=B_0$ then the designer is indifferent over all classifiers when $r^*=0$. In this case, the designer cares solely about the behavior of the individuals, and not how they are classified. As he can't affect their behavior when the reward is zero, all classifiers are optimal, including null ones. 

We'll move on to assume that $A_1\geq A_0$ and $B_1\geq B_0$, with one inequality strict. At a null equilibrium, $r^*=0$, and so conditional on prevalence $\pi_F=F(0)$ it must be optimal for $D$ to use a null classifier. If it was not optimal for $D$ to use a null classifier at $r^*=0$, then we could not support $r^*=0$ as an equilibrium, as the median voter strictly wants to either incentivize or disincentivize compliance (because $k_\mu^*(t,F)\not=0$).   

When $r^*=0$, $D$'s choice of classifier can't affect compliance, and so his problem is linear in $\delta_1, \delta_0$. Accordingly, letting
\begin{eqnarray*}
	\chi_1(\delta,\phi,r,F) & \equiv & \frac{\pi_F(\delta,\phi,r)\phi}{\pi_F(\delta,\phi,r)\phi+(1-\pi_F(\delta,\phi,r))(1-\phi)}, \text{ and }\\
	\chi_0(\delta,\phi,r,F) & \equiv & \frac{(1-\pi_F(\delta,\phi,r))\phi}{(1-\pi_F(\delta,\phi,r))\phi+\pi_F(\delta,\phi,r)(1-\phi)},
\end{eqnarray*}
then, conditional on a signal of $s_i=1$, $D$ will choose 
\[
d_i(s_i=1)=\begin{cases} 1 & \text { if }
	\chi_1(\delta,\phi,r,F)(A_1-B_0)+ B_0 \geq \chi_1(\delta,\phi,r,F)(A_0-B_1)+B_1,\\
	0 & \text{ otherwise.}
	% \chi_1(\delta,\phi,r,F)A_1+(1-\chi_1(\delta,\phi,r,F))B_0\geq \chi_1(\delta,\phi,r,F)A_0+(1-\chi_1(\delta,\phi,r,F))B_1,\\
	% 0 & \text{ otherwise.}
\end{cases}
\]
Similarly, conditional on a signal $s_i=0$, $D$ will choose 
\[
d_i(s_i=0)=\begin{cases} 0 & \text { if }
	\chi_0(\delta,\phi,r,F)
	(B_1-A_0)+ A_0 \geq \chi_0(\delta,\phi,r,F)(B_0-A_1)+A_1,\\
	1 & \text{ otherwise.}
	% \chi_0(\delta,\phi,r,F)
	% B_1+(1-\chi_0(\delta,\phi,r,F))A_0> \chi_0(\delta,\phi,r,F)B_0+(1-\chi_0(\delta,\phi,r,F))A_1,\\
	% 1 & \text{ otherwise.}
\end{cases} 
\]
It follows that when $F(0)\in \left( {{(B_1-B_0)(1-\phi)}\over{B_1-B_0-\phi(A_0-A_1+B_1-B_0)}},{{(B_1-B_0)\phi}\over{A_1-A_0-\phi(A_1-A_0-B_1+B_0)}}\right)$ it is optimal for $D$ to set $d_i=s_i$, and to choose $(\delta_1, \delta_0)=(1, 1)$. Consequently there can't be a null equilibrium in this case.  This is the only possible non-null classifier for $D$; the fact that we have assumed that $A_1\geq A_0$ and $B_1\geq B_0$ means that is is not possible for $d_i=1-s_i$ to be optimal for $D$. Consequently, if our condition on $F(0)$ doesn't hold, it must be the case that it is optimal for $D$ to choose a null classifier at $r^*=0$. And at any null classifier, $r^*=0$ is optimal for the median voter. 
\end{proof}

\noindent \textbf{Proposition \ref{theEquilibria}} 
\textit{In any non-null equilibrium, $r^*={k_\mu^*(t,F)\over{(\delta_1^*+\delta_0^*-1)(2\phi-1)}}$  and $\delta^*$ is as follows: }
\begin{itemize}
\item \textit{If $k_\mu^*(t,F)>0$}, 
\[
\delta^*=\begin{cases}
(1, \delta_0^*(k_\mu^*(t,F), 1)) \text{ with }\delta_0^*(k_\mu^*(t,F), 1)\not=0, \text{ or }\\
 (\delta_1^*(k_\mu^*(t,F), 0), 0) \text{ with }\delta_1^*(k_\mu^*(t,F), 0)\not=1
\end{cases}
\]
\item \textit{If $k_\mu^*(t,F)<0$, }
\[
\delta^*=\begin{cases}
(0, \delta_0^*(k_\mu^*(t,F), 0)) \text{ with }\delta_0^*(k_\mu^*(t,F), 0)\not=1,\text{ or }\\
 (\delta_1^*(k_\mu^*(t,F), 1), 1)\text{ with }\delta_1^*(k_\mu^*(t,F), 1)\not=0.
\end{cases}
\]
\end{itemize}
\noindent \textit{Proof:} The terms $\delta_0^*(k_\mu^*(t,F), 1), \delta_0^*(k_\mu^*(t,F), 0), \delta_1^*(k_\mu^*(t,F), 1)$, and $\delta_1^*(k_\mu^*(t,F), 0)$ are defined in Section  \ref{eqAppendix} of the Appendix, and represent the (unique and possibly interior) critical points of the designer's payoff function. 

Note that when $k_\mu^*(t,F)>0$ ($k_\mu^*(t,F)<0$) then the algorithm and reward must be positively (negatively) responsive in equilibrium. This means that more (less) individuals are being incentivized to engage in $\beta_i=1$ than would in the absence of any reward.  When $k_\mu^*(t,F)>0$, it must be the case that $r^*\cdot (\delta_1^*+\delta_0^*-1)>0$.  Consequently either $r^*>0$ and $\delta_1^*+\delta_0^*>1$, or $r^*<0$ and $\delta_1^*+\delta_0^*<1$.  In the former case, $D$'s payoffs are strictly quasiconcave in $\delta_0$ and strictly quasiconvex in $\delta_1$ (Proposition \ref{bigConcavity}).  Consequently, $\delta_1^*=1$ and $\delta_0^*\in\{\delta_0^*(k_\mu^*(t,F), 1), 1\}$.  In the latter case, $D$'s payoffs are strictly quasiconcave in $\delta_1$ and strictly quasiconvex in $\delta_0$. Consequently, $\delta_0^*=0$ and $\delta_1^*\in\{\delta_1^*(k_\mu^*(t,F), 0), 0\}$. The case of $k_\mu^*(t,F)<0$ can be proved similarly. $\Box$

\ \\ \textbf{Proposition \ref{null}} \textit{If $k_\mu^*(t,F)>0$  then there does not exist a non-null equilibrium when $A_1=B_0$, or when $B_1=B_0>A_1=A_0$. If $k_\mu^*(t,F)<0$  then there does not exist a non-null equilibrium when $A_0=B_1$, or when $A_1=A_0>B_1=B_0$.}
\begin{proof} The cases where $A_1=B_0$ or $B_1=A_0$ follow directly from evaluating the terms $\Delta_i^*(k, \delta_j)$ at $A_1=B_0$ and at $A_0=B_1$.    When $A_1=B_0$ and $k_\mu^*(t,F)>0$ then non-null equilibria must be of the form $(\delta_1, \delta_0)=(1, \delta^*_0(k_\mu^*(t,F), 1))$ or $(\delta_1, \delta_0)=(\delta^*_1(k_\mu^*(t,F), 0), 0)$.  However, when $k_\mu^*(t,F)>0$ then $\delta^*_0(k_\mu^*(t,F), 1)=0$ and $\delta^*_1(k_\mu^*(t,F), 0)=1$.  Consequently, there is not a non-null classifier that maximizes $D$'s payoff, as $D$ seeks to minimize $\delta^*_0(k_\mu^*(t,F), 1)$ over the set $(0, 1]$ and to maximize $\delta^*_1(k_\mu^*(t,F), 0)$ over the set $[0, 1)$.  The case of $B_1=A_0$ is proved similarly.

When $A_1=A_0>B_1=B_0$ or $B_1=B_0>A_1=A_0$  we can prove this result via Corollary \ref{Pr:OptimalComplianceAlgorithm}.  Suppose that $A_1=A_0>B_1=B_0$.  By our corollary, if $r>0$ then $(1, 1)$ is the unique optimal classifier for $D$ and if $r<0$ then $(0, 0)$ is the unique optimal classifier.  However, if $k_\mu^*(t,F)<0$ then $r^*(\delta^*_1+\delta^*_0-1)<0$ at a non-null equilibrium, which implies that either $r^*>0$ and $\delta^*_1+\delta^*_0-1<1$ or $r^*<0$ and $\delta^*_1+\delta^*_0-1>1$.  These inequalities are both inconsistent with optimizing behavior by $D$, and consequently there doesn't exist a non-null equilibrium.    
\end{proof}

\noindent \textbf{Proposition \ref{Pr:SocialWelfareOptimum}}
    \textit{For any precision $\phi \in \left(\frac{1}{2},1\right]$ and non-null classifier $\delta$, the social welfare maximizing reward, $r^*_{SW}(\delta,\phi)$ is defined by:}
    \[
        r^*_{SW}(\delta,\phi)=\frac{t}{\rho(\delta, \phi)}.    
    \]
\begin{proof}
The necessary first order condition for maximizing social welfare is: 
\begin{equation}
\label{swFOC}
f(r\cdot\rho(\delta, \phi))(-r\cdot \rho(\delta, \phi)^2 +t\cdot \rho(\delta, \phi) )=0.
\end{equation} 
Equation \eqref{swFOC} implies that social welfare is uniquely satisfied by 
\[
r^*_{SW}(\delta,\phi)={t\over{\rho(\delta, \phi)}}.
\]
We do not need to check the second order sufficient condition. This  must be a global maximum, as Equation \ref{swFOC} is positive for $r< {t\over{\rho(\delta, \phi)}}$ and negative for $r> {t\over{\rho(\delta, \phi)}}$.
\end{proof}

\normalsize 

\newpage
\bibliography{john-fairness}

\end{document}